\documentclass[aps, pra, twocolumn, longbibliography, superscriptaddress]{revtex4-1}

\usepackage[table]{xcolor}
\usepackage{physics}
\usepackage{amsmath, amssymb}
\usepackage{graphicx}
\usepackage{dcolumn}	
\usepackage{bm} 

\usepackage{mathtools}	

\usepackage[caption=false]{subfig}
\usepackage{verbatim}
\usepackage{hyperref}	

\begin{document}

\title{Hardware-Encoding Grid States in a Non-Reciprocal Superconducting Circuit}

\author{M. Rymarz}
\email{martin.rymarz@rwth-aachen.de}
\affiliation{JARA-Institute for Quantum Information, RWTH Aachen University, D-52056 Aachen, Germany}

\author{S. Bosco}
\affiliation{Peter Gr{\"u}nberg Institute, Theoretical Nanoelectronics, Forschungszentrum J{\"u}lich,\\ D-52425 J{\"u}lich, Germany}
\affiliation{Department of Physics, University of Basel, CH-4056 Basel, Switzerland}

\author{A. Ciani}
\affiliation{QuTech, Delft University of Technology, 2628 CJ Delft, The Netherlands}

\author{D. P. DiVincenzo}
\affiliation{JARA-Institute for Quantum Information, RWTH Aachen University, D-52056 Aachen, Germany}
\affiliation{Peter Gr{\"u}nberg Institute, Theoretical Nanoelectronics, Forschungszentrum J{\"u}lich,\\ D-52425 J{\"u}lich, Germany}

\begin{abstract}
We present a circuit design composed of a non-reciprocal device and Josephson junctions whose ground space is doubly degenerate and the ground states are approximate codewords of the Gottesman-Kitaev-Preskill (GKP) code.
We determine the low-energy dynamics of the circuit by working out the equivalence of this system to the problem of a single electron confined in a two-dimensional plane and under the effect of strong magnetic field and of a periodic potential.
We find that the circuit is naturally protected against the common noise channels in superconducting circuits, such as charge and flux noise, implying that it can be used for passive quantum error correction.
We also propose realistic design parameters for an experimental realization and we describe possible protocols to perform logical one- and two-qubit gates, state preparation and readout.
\end{abstract}

\maketitle

\section{Introduction}
Building a quantum computer in a physical system is a formidably challenging task because of the inherent fragility of physical quantum bits (qubits). The key idea behind quantum error correction (QEC) \cite{Shor, Terhal} is to use \textit{logical} qubits that can be protected against certain likely errors, thus extending the lifetime of the encoded quantum information and allowing for fault tolerant quantum computation \cite{Preskill, Chao}.

There are different flavors of QEC codes, that differ in the way in which the logical qubits are constructed. For example, in the toric \cite{Kitaev_toric, Dennis}, surface \cite{Bravyi_code, Freedman, Wallraff} and color \cite{Bombin} code, the logical qubits are encoded in lattices of physical qubits. To date, the QEC codes that have been most successful in enhancing the lifetime of quantum information have been built from continuous variable (CV) systems \cite{Braunstein, Slotine, Lloyd}, such as a single microwave cavity mode. Efficient QEC with cat-states and binomial codes has been demonstrated \cite{Ofek, Hu}. In this work, we focus on a similar CV encoding, proposed by Gottesman, Kitaev and Preskill (GKP) in Ref. \cite{GKP}, where the codewords are shifted grid states and can be protected against sufficiently small translations in phase space. The error correcting properties of the GKP code have been further explored in Ref. \cite{Albert}, where it was shown that the GKP code outperforms cat and binomial codes when a photon loss channel is considered. Grid states have been successfully prepared and actively stabilized in superconducting cavities by a stroboscopic modulation of interactions \cite{Eickbusch}. Also, passive implementations of these states in superconducting circuits have been proposed with the $0$-$\pi$ qubit \cite{Kitaev_protected, BKP, Dempster, Groszkowski, Gyenis, Paolo, Smith} and the dualmon \cite{Grimsmo}.

However, the \textit{active} implementation of QEC requires complicated protocols where errors are detected and compensated for by applying a recovery operation. In contrast, in \textit{passive} QEC the protection is a built-in feature of the system's hardware and it is therefore advantageous in terms of hardware efficiency and scalability. Generally, this is achieved by constructing a system whose two-fold degenerate ground states are the qubit states: errors that bring the system out of the computational space have an associated energy penalty and so the system will automatically relax back into the computational space \cite{Kitaev_toric, Doucot_protected, bravyiHastings2010}.

An example of an implementation of the GKP code is a single electron confined in a two-dimensional plane within a periodic potential and a high perpendicular magnetic field \cite{Onsager, Harper, Azbel, Zak4, Zak2, Zak3, Zak, Rauh_Wannier, Rauh1, Rauh2, Hofstadter, Falicov, Wannier, Geisel}. 
The magnetic field restricts the dynamics of the electron to the lowest Landau level (LLL), so that the position operators in orthogonal directions do not commute. The contribution of the periodic potential to the Hamiltonian reduces to a sum of displacement operators, which are the stabilizers of the GKP code.

Although this system is useful for a theoretical understanding of the code, it is very unpractical to implement. The magnetic field required for it to work is exactly $B=\Phi_0/2A$ with $A$ the area of the unit cell in space and the magnetic flux quantum $\Phi_0=h/e$. In realistic crystals, this condition implies that the magnetic field needs to be unrealistically large, about $B \sim 10^5\,\text{T}$. Moir\'{e} patterns in twisted bilayer graphene can be used to reduce this value by a few orders of magnitude due to their large unit cell \cite{Dean}. Even if such a regime were possible to achieve, the external magnetic field would still require an extremely precise fine-tuning and the electron density would still have to be decreased to the unfeasible value of a single free electron in the crystal. \\

\begin{figure}[t!]
	\centering
	\includegraphics[width=0.40\textwidth]{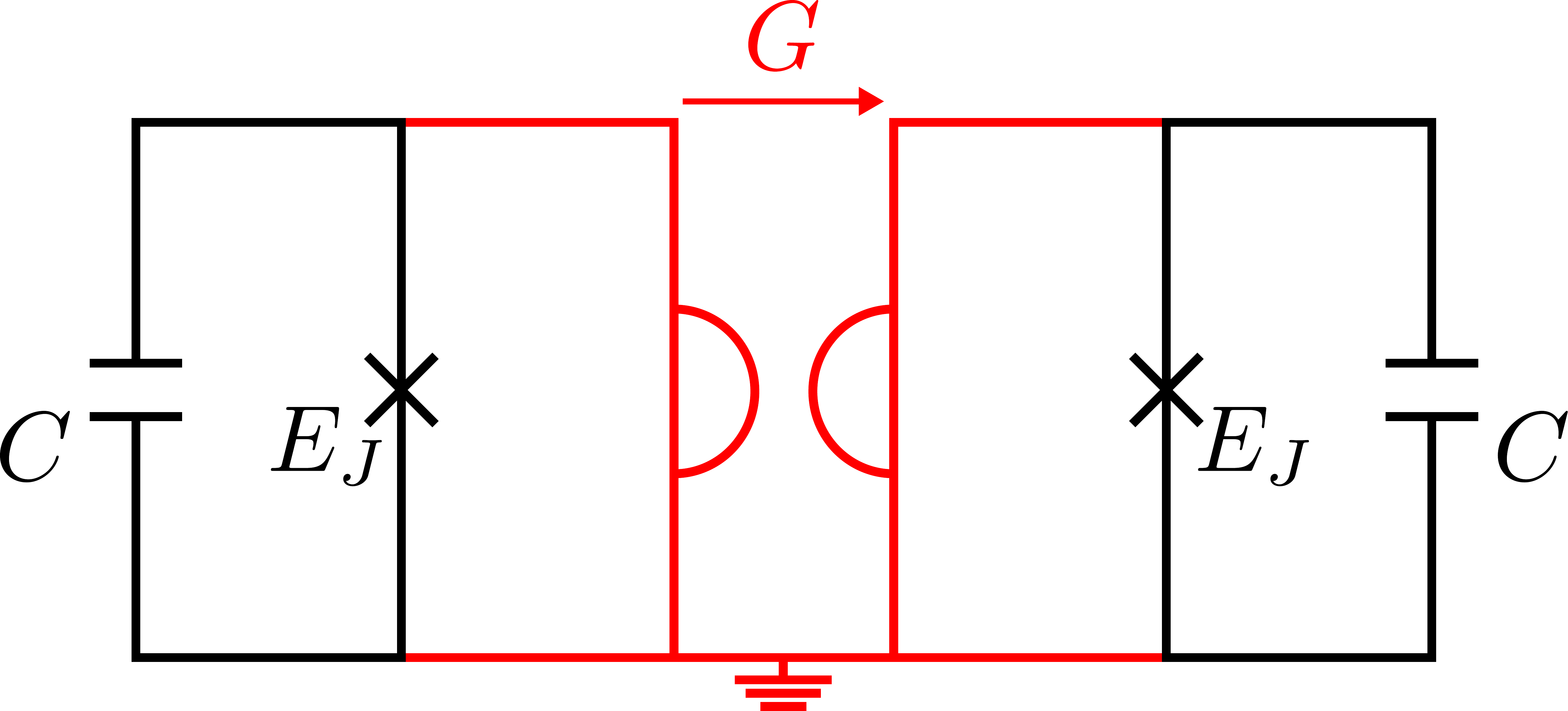}
	\caption{Proposed hardware implementation of the GKP code. The circuit consists of two Josephson junctions coupled by a gyrator, highlighted in red.}
	\label{fig_gyrator_introduction}
\end{figure}

Here, we propose instead a different implementation of the GKP code, which does not suffer from any of these issues. We consider a superconducting circuit composed of two Josephson junctions coupled by a classic, lossless, linear non-reciprocal circuit element, the \textit{gyrator} \cite{Tellegen}. The circuit is shown in Fig. \ref{fig_gyrator_introduction}.

\newpage
The non-reciprocity of the gyrator breaks time-reversal symmetry and its contribution to the dynamics of the circuit is akin to that of a uniform, perpendicular magnetic field in an electronic system. In our circuit, the condition on the strength of the magnetic field reduces to the requirement on the gyration conductance being precisely twice the conductance quantum, i.e. $G=2e^2/h$.

While unrealistic for conventional gyrators \cite{Hogan1, Hogan2, Rosenthal, Chapman, Lecocq, barzanjeh2017}, this value of $G$ can be easily reached in quantum Hall effect (QHE) devices \cite{Viola, Bosco1, Bosco2, BoscoReilly, Mahoney}, in which a precise fine-tuning of the device parameters is unnecessary due to the quantization of the off-diagonal conductivity. In addition, although conventional QHE devices \cite{Mahoney} require a high magnetic field to operate, making them unpractical to couple to superconducting devices, we note that state-of-the-art quantum anomalous Hall effect materials still present an extremely precise conductivity quantization and low losses \cite{Bestwick} and can be used to implement non-reciprocal electrical network elements that can operate at zero magnetic field \cite{Reilly}.

We show that our construction is insensitive to common types of noise. We discuss possible ideas of how logical one-qubit as well as two-qubit Clifford gates can be implemented by applying currents and using tunable inductances. We show that the ground state of our system is an eigenstate of the Hadamard gate. Consequently, our system is suitable for universal quantum computation \cite{GKP, BravyiKitaev, Yamasaki}. \\

The paper is organized as follows:
In Sec. \ref{sec_GKP} we review a few key concepts of hardware-encoding GKP states and we introduce the Hamiltonian whose two-fold degenerate ground space is spanned by the GKP codewords.
In Sec. \ref{sec_crystal} we show how this Hamiltonian can be derived from the low-energy description of a single electron in a high magnetic field and a periodic potential. The dynamics of this system is equivalent to that of a gyrator connected to two Josephson junctions but the solid-state jargon more easily reveals the intimate relation to Hoftstader's butterfly \cite{Hofstadter, Geisel}: the GKP states are obtained at a specific point in the butterfly. 
In Sec. \ref{sec_confinement} we study the effect of an additional parabolic confinement potential, which in the circuit model consists of the addition of inductances in parallel to the Josephson junctions. For this setting, the ground space of the resulting Hamiltonian is two-fold degenerate up to an exponentially small gap and the eigenstates of the system resemble superpositions of normalizable, approximate GKP codewords \cite{GKP}.
In Sec. \ref{sec_full_wf} we highlight the connection between the one-dimensional GKP grid states and the two-dimensional ground space wave functions of the electronic system's Hamiltonian projected onto the lowest Landau level.
In Sec. \ref{sec_circuit} we work out in detail the equivalence to the non-reciprocal superconducting circuit model and we propose realistic design parameters for an experimental realization of the system. We also discuss possible realizations of logical gates by using current sources and tunable inductances and we present ideas for state preparation and readout. We provide an analysis of the protection against common noise sources, such as flux and charge noise.
Finally, in Sec. \ref{sec_conclusions} we summarize our results and give an outlook on further work.

\section{The GKP code for passive QEC}\label{sec_GKP}
The GKP code is a CV quantum error correcting code \cite{Braunstein, Slotine, Lloyd} introduced in Ref. \cite{GKP}. In contrast to the standard approach to quantum error correction, which assumes physical qubits as fundamental noisy elements, in CV quantum error correction the idea is to encode a two-level (or $n$-level more generally) system in the infinite-dimensional Hilbert space of an one-dimensional particle characterized by dimensionless canonical quadrature operators $X$ and $P$ satisfying $[X, P]=i$. The GKP code can be described within the stabilizer formalism for CV systems. The role of the Pauli group is played by the Weyl-Heisenberg group $G_{\rm H}$, i.e. the group of displacement operators \cite{Weedbrook, Gerry} 
\begin{equation}\label{eq_displacement_operator}
D(\alpha)=e^{\alpha b^\dagger - \alpha^* b},	\qquad \alpha \in \mathbb{C},
\end{equation}
with the annihilation operator $b = (X+i P)/\sqrt{2}$. In  this framework the GKP code is the $2$-dimensional subspace stabilized by a subgroup $\mathcal{S}_{\rm GKP}$ of $G_{\rm H}$ with group generators 
\begin{equation}\label{eq_stabilizer}
	S_{X} = D ( i \sqrt{2 \pi} ) = e^{i 2 \sqrt{\pi}  X}, \quad 
	S_{P} = D ( \sqrt{2 \pi} ) = e^{-i 2 \sqrt{\pi} P}
\end{equation}
and $S_{X}^{-1}, S_{P}^{-1}$. The logical Pauli operators $\overline{Z}$ and $\overline{X}$ are given by
\begin{equation}\label{eq_logical_X_Z}
	\overline{Z} = S_{X}^{1/2}= e^{ i \sqrt{\pi}  X}, \qquad 
	\overline{X} = S_{P}^{1/2}=e^{-i \sqrt{\pi} P}.
\end{equation}
This choice of logical operators fixes the following (unnormalizable) codewords
\begin{subequations}\label{eq_GKP_code_words}
\begin{align}
	\ket{\bar{0}} &= \sum_{n \in \mathbb{Z}} \ket{X =2\sqrt{\pi}n},						\\
	\ket{\bar{1}} &= \sum_{n \in \mathbb{Z}} \ket{X = 2\sqrt{\pi}n + \sqrt{\pi}},
\end{align}
\end{subequations}
that are grid states, each describing a comb of equidistant $\delta$-peaks in the $X$-basis. These combs have a period of $2 \sqrt{\pi}$ and are shifted with respect to each other by $\sqrt{\pi}$.

Given a density matrix $\rho$ describing the state of an one-dimensional CV quantum system, one can expand a generic quantum operation $\mathcal{E}(\rho)$ in terms of displacement operators as \cite{GKP}
\begin{equation}
	\mathcal{E}(\rho)= \int_{\mathbb{C}} d \alpha \int_{\mathbb{C}} d \beta f(\alpha, \beta) D(\alpha) \rho  D^{\dagger}(\beta), 
\end{equation}
with $f(\alpha, \beta)$ being a scalar function. If the function $f(\alpha, \beta)$ has support only on a domain in which $D^{\dagger}(\beta)D(\alpha)$ is either in the stabilizer group or does not commute with the stabilizers $S_{X}$ and $S_{P}$ in Eq. \eqref{eq_stabilizer}, then the GKP code can correct against these kinds of errors, provided that shifts in position and momentum obey
\begin{equation}
	\lvert \Delta_X \rvert < \frac{\sqrt{\pi}}{2}, \qquad 
	\lvert \Delta_P \rvert <\frac{\sqrt{\pi}}{2}. 
\end{equation}
In this case, the error syndromes are unique. Otherwise, logical errors will be made.

The main idea behind passive, stabilizer error correction is to construct a Hamiltonian that has the code subspace as the low energy subspace. For the GKP code, this Hamiltonian is easily obtained as \cite{GKP}
\begin{equation}\label{eq_GKP_Hamiltonian}
	H_\text{GKP}/V_0 = 
	-\bigl[\cos \bigl(2 \sqrt{\pi} X \bigr) 
	+ \cos \bigl(2 \sqrt{\pi} P\bigr) \bigr],
\end{equation}
with $V_0$ a constant with the unit of energy. Because the code subspace is stabilized by the two cosines it has energy $-2 V_0$.

We remark that passive, stabilizer error correction for CV systems is rather different than in systems based on a large set of physical qubits, such as the toric, surface or color code \cite{Kitaev_toric, Dennis, Bravyi_code, Freedman, Bombin}. In fact, the Hamiltonian in Eq. \eqref{eq_GKP_Hamiltonian} is gapless, with a continuous spectrum ranging from $-2 V_0$ to $+2 V_0$. Also, because the Weyl-Heisenberg group is a continuous group, in contrast
to the discrete Pauli group, the eigenstates of $H_\text{GKP}$ are unnormalizable and formally out of the Hilbert space of any physical system. As a consequence, the usual perturbation theory argument \cite{Kitaev_toric, bravyiHastings2010} which claims that $\textit{local}$ perturbations of the Hamiltonian give rise to small variations of the energy levels is not applicable here. As pointed out in Ref. \cite{Doucot_protected}, the argument can be restored in approximated versions of Eq. \eqref{eq_GKP_Hamiltonian}, where the eigenstates are normalized and confined, leading to a discrete spectrum. The particular way in which $H_\text{GKP}$ is approximated modifies the properties of the degenerate ground space, but generally its eigenstates remain with disjoint support. This idea of passive protection is similar to the one behind the $0$-$\pi$ qubit \cite{Kitaev_protected, BKP, Dempster, Groszkowski, Gyenis, Paolo, Smith}.

\section{Crystal Electron in Magnetic Field}\label{sec_crystal}
We discuss how the code Hamiltonian in Eq. \eqref{eq_GKP_Hamiltonian} can emerge from the consideration of the well-known situation of a single electron confined to a two-dimensional plane in a strong perpendicular uniform magnetic field \cite{Onsager, Harper, Azbel, Zak4, Zak2, Zak3, Zak, Rauh_Wannier, Rauh1, Rauh2, Hofstadter, Falicov, Wannier, Geisel}. The effect of a periodic potential on the electron's motion has been extensively studied because of the fractal nature of the energy bands \cite{Hofstadter} and their non-trivial topology \cite{Thouless}. In this section, we want to clarify under what conditions the ground states of the system are GKP states. We focus on the Hamiltonian
\begin{equation}\label{eq_crystal_Hamiltonian}
	H = \frac{\left[ \bm{p}+e\bm{A}(x_1,x_2) \right]^2}{2m} + V_\text{crys}(x_1, x_2),
\end{equation}
where the two-dimensional positions and momenta satisfy canonical commutation relations $[x_i,p_j]=i\hbar\delta_{ij}$, for $i,j \in \{1,2\}$, and $\bm{B} = \grad \cross \bm{A} = B \bm{e}_3$. We consider a  crystal potential of the form
\begin{equation}\label{eq_V_crys}
	V_\text{crys}(x_1,x_2) = - V \left[ \cos(2\pi \frac{x_1}{L_0}) + \cos(2\pi \frac{x_2}{L_0}) \right],
\end{equation}
which corresponds to the first Fourier mode of any periodic potential on a square lattice of size $L_0$ in the $x_1x_2$-plane. Although both the crystal potential $V_\text{crys}$ and the uniform magnetic field $\bm{B}$ are periodic in the $x_1$- and $x_2$-direction, the Hamiltonian is not, because the discrete translation symmetry is broken in at least one direction by the vector potential $\bm{A}$. As a result, $H$ does not simultaneously commute with both the canonical unitary translation operators $t_1(L_0)$ and $t_2(L_0)$, defined as
\begin{equation}\label{eq_def_TO}
	t_i(r) = e^{-irp_i/\hbar},	\qquad r \in \mathbb{R}, \qquad i =1,2.
\end{equation}
It follows that $H, t_1(L_0)$ and $t_2(L_0)$ cannot have common eigenstates, as the usual formulation of Bloch's theorem dictates. To find a set of translations which do commute with the Hamiltonian, we work with the dynamical momenta \cite{Girvin_Prange, Doucot_magnet, Tong, Girvin}
\begin{equation}\label{eq_def_dynamical_momenta}
	\pi_1 = p_1 + eA_1,		\qquad
	\pi_2 = p_2 + eA_2,
\end{equation}
and the guiding center variables
\begin{equation}\label{eq_def_guiding_center}
	R_1 = x_1 - \frac{1}{m\omega_c} \pi_2,	\qquad
	R_2 = x_2 + \frac{1}{m\omega_c} \pi_1,
\end{equation}
with the cyclotron frequency $\omega_c = eB/m$. These operators are gauge-invariant (in contrast to the canonical momenta $p_i$) and satisfy the commutation relations
\begin{equation}
 [\pi_1, \pi_2] = -i \frac{\hbar^2}{l^2_B},
 \qquad
 [R_1, R_2] =  i l_B^2,
 \qquad
 [\pi_i, R_j] = 0,
\end{equation}
for $i,j \in \{1,2\}$ and the magnetic length $l_B = \sqrt{\hbar/eB}$. 
Physically, the dynamical momenta are related to the cyclotron motion of an electron around its center of mass, which in turn is parametrized by the guiding center coordinates.

We can impose boundary conditions by requiring the wave function to be quasi-periodic in $R_i$. To this end, we make use of the unitary \textit{magnetic} translation operators (MTOs) \cite{Zak4, Girvin, Brown, Fischbeck} \footnote{In the $x_i$-representation, the MTOs differ from the conventional translation operators, as defined in Eq. \eqref{eq_def_TO}, by an additional $x_i$-dependent complex phase, originating from the gauge of the vector potential $\bm{A}(x_1,x_2)$.}
\begin{equation}\label{eq_MTO_def}
	T_1(r) = e^{-i r R_2 /l_B^2}
	,	\qquad
	T_2(r) = e^{i r R_1 /l_B^2}
	,	\qquad r \in \mathbb{R},
\end{equation}
which shift the guiding center variables $R_i$ by $r$, i.e.
\begin{equation}\label{eq_action_of_MTO}
	T_i^\dagger(r) R_i T_i(r) = R_i + r, \qquad i=1,2.
\end{equation}
It is straightforward to show that the MTOs $T_1(L_0)$ and $T_2(L_0)$ do commute with the Hamiltonian in Eq. \eqref{eq_crystal_Hamiltonian}. However, because of the non-commutativity of $R_1$ and $R_2$ in Eq. \eqref{eq_def_guiding_center}, magnetic translations in different directions do not generally commute. An electron moving along a closed path accumulates an Aharonov-Bohm phase \cite{AharonovBohm} proportional to the magnetic flux threaded by the loop and so
\begin{equation}\label{eq_MTO_commutation}
	T_2(r_2) T_1(r_1) = e^{i 2 \pi B r_1 r_2 / \Phi_0} T_1(r_1) T_2(r_2),
\end{equation}
with the (non-superconducting) flux quantum $\Phi_0=h/e$. Consequently, MTOs in orthogonal directions commute only when an integer number of flux quanta is threaded through the loop defined by the MTOs. Note that with an appropriate rescaling, the MTOs defined in Eq. \eqref{eq_MTO_def} correspond to the displacement operators similar to the ones defined in Eq. \eqref{eq_displacement_operator}.

In the following, we restrict to \textit{rational} fluxes \cite{Hofstadter}, where the magnetic flux enclosed in a unit cell of size $L_0 \times L_0$ is a rational multiple of the flux quantum, i.e.
\begin{equation}\label{eq_def_fraction}
	\Phi = B L_0^2 = \frac{p}{q} \Phi_0,
\end{equation}
with coprime natural numbers $p$ and $q$. In this case, we consider an enlarged, \textit{magnetic} unit cell of size $q L_0 \times L_0$, which contains $p$ flux quanta, such that the MTOs $T_1(qL_0) = \left[ T_1(L_0) \right]^q$ and $T_2(L_0)$ commute with the Hamiltonian in Eq. \eqref{eq_crystal_Hamiltonian} and with each other \footnote{We remark that the direction of the enlargement is arbitrary.}. As a result, we consider the magnetic Bloch states satisfying
\begin{subequations}\label{eq_toroidal_BC}
\begin{align}
	T_1(qL_0) \ket{\bm{k}} &= e^{i k_1 q L_0} \ket{\bm{k}},	\\
	T_2(L_0) \ket{\bm{k}} &= e^{i k_2 L_0} \ket{\bm{k}},
\end{align}
\end{subequations}
where $\bm{k}=(k_1, k_2)^T$ is the crystal momentum defined in the rectangular Brillouin zone 
\begin{equation}\label{eq_Brillouin_zone}
	k_1 \in \big[ 0, \tfrac{2\pi}{qL_0} \big),	\qquad
	k_2 \in \big[ 0, \tfrac{2\pi}{L_0} \big).
\end{equation}
The states $\ket{\bm{k}}$ are sometimes referred to as Zak states \cite{Zak, Zak2, Zak3}.

Introducing the Landau level ladder operators 
\begin{equation}\label{eq_ladder_momenta}
	a =  \frac{1}{\sqrt{2}}\frac{l_B}{\hbar}(\pi_2 + i \pi_1),		\qquad
	a^\dagger =  \frac{1}{\sqrt{2}}\frac{l_B}{\hbar}(\pi_2 - i \pi_1),
\end{equation}
satisfying the bosonic commutation relation $[a,a^\dagger]=1$, the Hamiltonian in Eq. \eqref{eq_crystal_Hamiltonian} can be rewritten as
\begin{equation}\label{eq_final_crystal_Hamiltonian}
\begin{split}	
	H = \hbar \omega_c &\bigg( a^\dagger a + \frac{1}{2} \bigg) - \frac{V}{2} \bigg[ D_a \left( \tfrac{i\sqrt{2}\pi l_B}{L_0} \right) T_1 \left( \tfrac{qL_0}{p} \right)		\\	
	&+ D_a \left( \tfrac{-\sqrt{2}\pi l_B}{L_0} \right) T_2 \left( \tfrac{q L_0}{p} \right) + \mathrm{h.c.} \bigg],
\end{split}
\end{equation}
with the cyclotron frequency $\omega_c = eB/m$ and the unitary displacement operator $D_a(\alpha)=\exp\left(\alpha a^\dagger - \alpha^* a \right)$ acting on the subspace of the dynamical momenta.
A convenient basis to numerically analyze the low energy spectrum of this Hamiltonian are the product states $\ket{n;\bm{k},l} = \ket{n} \otimes \ket{\bm{k},l}$ satisfying
\begin{equation}\label{eq_def_Landau_Level_Fock_state}
	a^\dagger a \ket{n} = n \ket{n}, 	\qquad n \in \mathbb{N}_0,
\end{equation}
and
\begin{subequations}\label{eq_toroidal_2}
\begin{align}
	T_1(qL_0) \ket{\bm{k},l} &= e^{i k_1 q L_0} \ket{\bm{k},l},	\\
	T_2(L_0/p) \ket{\bm{k}, l} &= e^{i (k_2 L_0 + 2 \pi l)/p} \ket{\bm{k}, l}, \\
	\braket{\bm{k}, l}{\bm{k}, l'} &\propto \delta_{ll'},
\end{align}
\end{subequations}
with $l=0,1,2,\ldots,p-1$, see Appendix \ref{app_MTO}.
Note that in the absence of the potential ($V=0$), the states $\ket{n;\bm{k},l}$ diagonalize the Hamiltonian, leading to the $p$-fold degenerate Landau level spectrum $E_n=\hbar \omega_c (n+1/2)$. Expressed in the $x_1x_2$-representation, the quasi-periodic wave functions $\Psi_{n;\bm{k},l}(x_1,x_2) = \braket{x_1,x_2}{n;\bm{k},l}$ were introduced by Haldane and Rezayi in Ref. \citep{Rezayi}, see also Sec. \ref{sec_full_wf}.

\begin{figure}[t]
	\centering
	\vspace{-10pt}
	\includegraphics[width=0.5\textwidth]{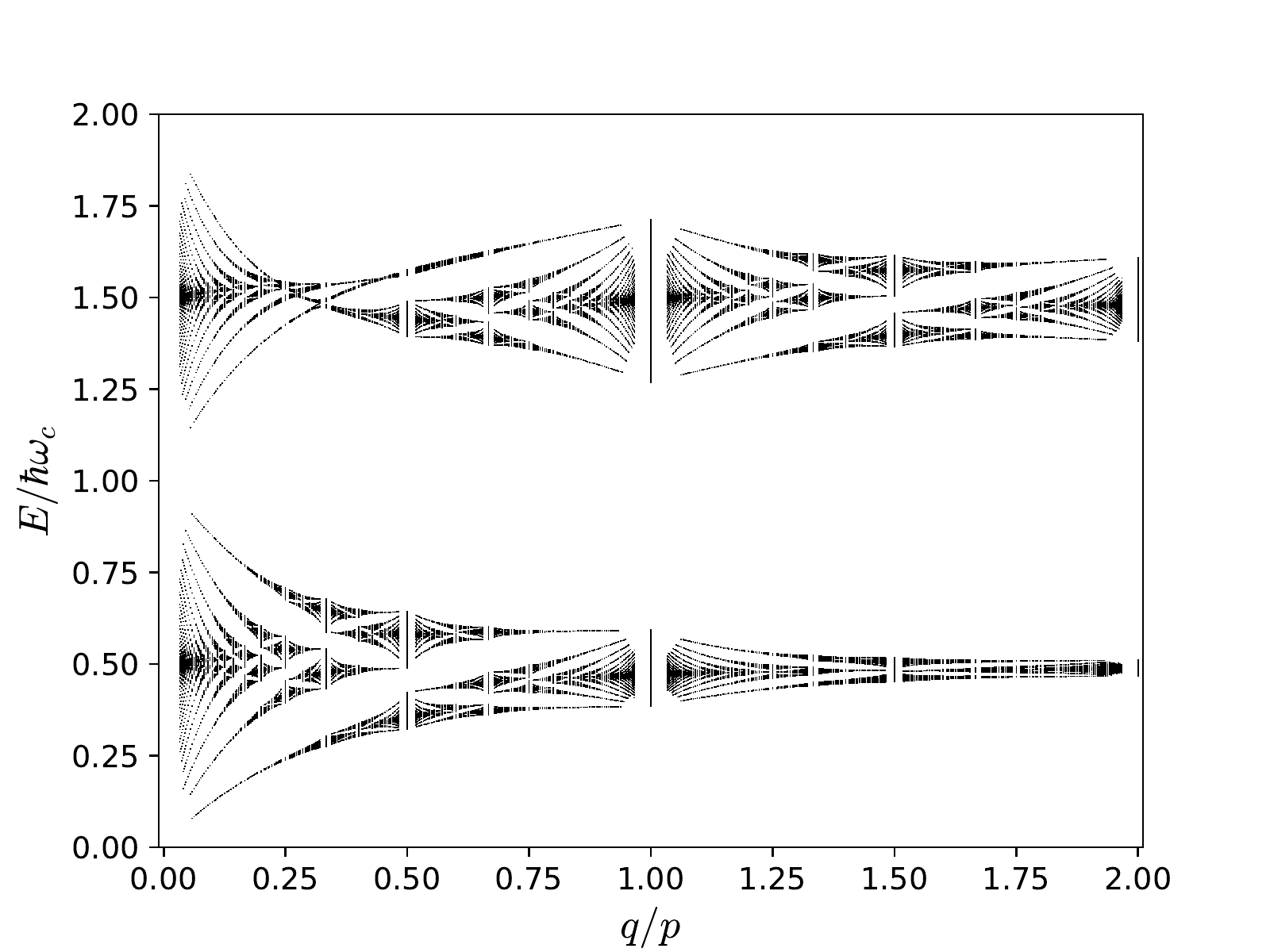}
	\caption{Low energy spectrum of the Hamiltonian in Eq. \eqref{eq_final_crystal_Hamiltonian} as a function of the \textit{inverse} flux ratio $q/p$ for a fixed value of $V/\hbar \omega_c = 0.25$. The initially flat Landau levels at $E_n/\hbar \omega_c = n+1/2$ (obtained for $V=0$) are split into subbands.}
	\label{fig_energy_spectrum}
\end{figure}

The crystal potential couples states with different Landau level occupation number $n$ and with different guiding center quantum number $l$. Consequently, the $p$-fold degeneracy is lifted and, for a moderately weak crystal potential, each Landau level splits into $p$ subbands with finite broadening and $q$-fold degeneracy \cite{Rauh1, Rauh2, Rauh_Wannier}, see Fig. \ref{fig_energy_spectrum}, in which the two lowest split Landau levels are shown. More details on the general solution of the Hamiltonian in Eq. \eqref{eq_final_crystal_Hamiltonian} can be found in Appendix \ref{app_numerics_crystal_electron}.

In this paper, we are interested in the weak Landau level coupling limit $V/\hbar\omega_c \ll 1$, where the dynamics of states within each Landau level can be taken to be independent from the others. This limit will be analyzed in the following.

\subsection{GKP Qubit in the LLL Projection}\label{subsection_LLL_projection}\label{subec_LLL_projection}
When the coupling between the Landau levels is weak, an effective low-energy Hamiltonian acting on a single Landau level can be obtained by a Schrieffer-Wolff transformation \cite{Winkler, Bravyi}.
In particular, to the lowest order and considering only the LLL \footnote{The treatment of higher Landau levels is similar and straightforward.}, we obtain from Eq. \eqref{eq_final_crystal_Hamiltonian} the effective Hamiltonian (up to an unimportant constant)
\begin{equation}\label{eq_Harper}
\begin{split}
	H_\text{LLL} 
	&= \bra{n=0} H \ket{n=0}		\\
	&= - \frac{V_0}{2}
	\bigg[ 
	T_1 \left( \tfrac{q L_0}{p} \right) + T_2 \left( \tfrac{q L_0}{p} \right) 
	+ \mathrm{h.c.} \bigg],
\end{split}
\end{equation}
where $V_0 = V e^{-\pi q / 2p}$. Although formally this effective Hamiltonian is valid only when $V/\hbar\omega_c \ll 1$, numerics shows that the approximation holds up well to relatively high values of $V/\hbar\omega_c  \lesssim 0.4$, see Sec. \ref{subsec_low_energy} where we discuss in more detail the validity of the LLL projection. 

\begin{figure}[t!]
	\centering
	\vspace{-15pt}
	\includegraphics[width=0.5\textwidth]{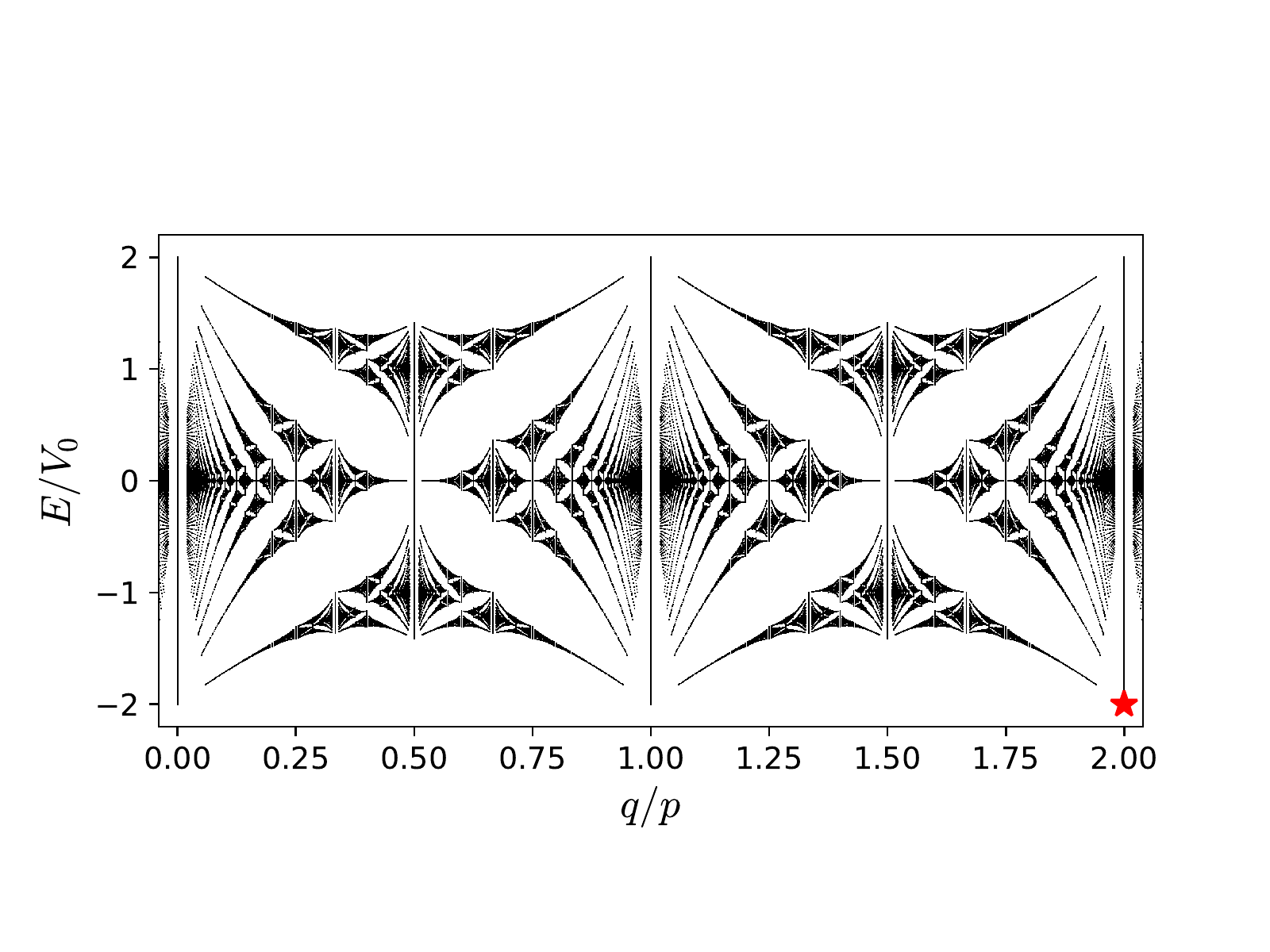}
	\vspace{-30pt}
	\caption{Hofstadter's butterfly obtained by plotting the spectrum of the effective lowest Landau level Hamiltonian $H_\text{LLL}/V_0$, defined in Eq. \eqref{eq_Harper}, as function of the inverse magnetic flux ratio $\Phi_0/\Phi=q/p$. The two-dimensional GKP code space corresponds to the states of minimal energy at $p/q=1/2$. This point is marked by a red star in the figure.}
	\label{fig_Hofstadter_butterfly}
\end{figure}

In the limit of weak Landau level coupling, the eigen equation associated with $H_\text{LLL}$ is the Harper equation \cite{Harper, Geisel, Azbel, Falicov, Bernevig}, which is a special case of the almost Mathieu equation \citep{Bellissard, Avila, Avila2}. In particular, the Harper equation is a finite-difference equation, resulting in an energy spectrum in form of the Hofstadter butterfly \cite{Hofstadter}, shown in Fig. \ref{fig_Hofstadter_butterfly}.

We point out that this spectrum has $p$ bands that are $q$-fold degenerate. In fact, states connected by the application of $T_1(nL_0)$ with $n=1, \ldots ,q-1$ are orthogonal but have the same energy, see Appendix \ref{app_MTO}. Note that this result is in contrast to the original tight-binding result of Hofstadter \cite{Hofstadter}, where different Landau levels are strongly coupled and where there are $q$ bands that are $q$-fold degenerate \cite{Bernevig}. For this reason, in the original work the Hofstadter butterfly is obtained by plotting the spectrum as a function of $p/q$ instead of $q/p$ \cite{Hofstadter}.

Importantly, by introducing the dimensionless variables
\begin{equation}\label{eq_X_P_rescaling}
	X = \frac{\sqrt{\pi}}{L_0} R_1 = \sqrt{\frac{q}{2p}} \frac{R_1}{l_B}, \qquad	
	P = \frac{2 p \sqrt{\pi}}{q L_0} R_2 = \sqrt{\frac{2p}{q}} \frac{R_2}{l_B},
\end{equation}
satisfying the canonical commutation relation $[X,P] = i$, we can rewrite $H_\text{LLL}$ as 
\begin{equation}\label{eq_eff_GKP_Hamiltonian}
	H_\text{LLL}/V_0
	=  - \bigg[ \cos(2 \sqrt{\pi} X) + \cos( \frac{q}{p} \sqrt{\pi} P) \bigg].
\end{equation}
Comparing with Eq. \eqref{eq_GKP_Hamiltonian}, we observe that $H_\text{LLL}$ corresponds to the GKP Hamiltonian $H_\text{GKP}$ when $p/q = 1/2$. This system is therefore suitable for passively encoding the GKP codewords, which are given in Eq. \eqref{eq_GKP_code_words}. Note that for $p/q=1/2$, the previous rescaling of the guiding center variables $R_i$ becomes equal, and so the code can correct equal shifts on $X$ and $P$. 

Furthermore, for $p/q=1/2$, the MTOs defining the states $\ket{\bm{k}}$ in Eq. \eqref{eq_toroidal_BC} are related to the stabilizers and logical operators of the GKP code as
\begin{equation}
	T_1(2L_0) = S_P,		\qquad
	T_2(L_0) = S_X^{1/2} = \overline{Z}.
\end{equation}
Since the GKP codewords are the eigenstates of $H_\text{LLL}$ with minimal eigenenergy, we can identify the code space with a specific point in Hofstadter's butterfly (see the red star in Fig. \ref{fig_Hofstadter_butterfly}). In particular, the code space is spanned by the eigenstates obtained for $\bm{k}=(0,0)^T$ and $\bm{k}=(0,\pi/L_0)^T$. These states correspond to the logical codewords introduced in Eq. \eqref{eq_GKP_code_words}. The eigenfunctions of the full system within the LLL projection will be analyzed in Sec. \ref{sec_full_wf}. \\

At this point, we would like to highlight the main difference between our approach and the original proposal by GKP \cite{GKP}. GKP proposed to use the LLL projection at the rational flux $p/q=d/1$ without the crystal potential ($V_\text{crys} \equiv 0$). A qudit can be encoded by focusing on the $d$-fold degenerate ground space obtained at vanishing Bloch momentum ($\bm{k}=\bm{0}$), and one can take this qudit to construct different shift resistant quantum codes. In contrast, for the rational flux $p/q=1/2$, by including the crystal potential and using states with different Bloch momenta, here we encode a qubit in a real CV system.

\section{Additional parabolic confinement}\label{sec_confinement}
Because the GKP codewords in Eq. \eqref{eq_GKP_code_words} are not normalizable, they are mathematical objects that are not physically realizable. Furthermore, the continuous spectrum of $H_\text{GKP}$ is problematic for the implementation of the GKP code in realistic systems, since noise or temperature would affect the states \cite{Doucot_protected}. For these reasons, we consider an additional parabolic potential in the Hamiltonian in Eq. \eqref{eq_crystal_Hamiltonian}. This potential renders the spectrum discrete and the states normalizable. We show that the eigenstates of this modified Hamiltonian are related to the approximate grid states introduced in the original work by GKP \cite{GKP}.

For simplicity, we choose the parabolic confining potential to be isotropic and so the Hamiltonian is
\begin{equation}\label{eq_confined_Hamiltonian_initial}
H = \frac{[\bm{p} + e\bm{A}(x_1, x_2)]^2}{2m} + V_\text{tot}(x_1,x_2),
\end{equation}
with
\begin{equation}\label{eq_total_potential}
	V_\text{tot}(x_1,x_2) = V_\text{crys}(x_1, x_2) + \frac{1}{2} m \omega_0^2 \left( x_1^2+x_2^2 \right).
\end{equation}
Because $V_\text{tot}$ does not preserve the discrete translational symmetry defined by $V_\text{crys}$, we cannot impose the magnetic Bloch conditions in Eq. \eqref{eq_toroidal_BC}. In this case, we require instead that the wave functions vanish at infinity.

We now briefly summarize the main findings of a numerical analysis of the eigensystem of the Hamiltonian in Eq. \eqref{eq_confined_Hamiltonian_initial} (see Appendix \ref{app_numerical_preparation} for details). In particular, we focus on the case  $p/q=1/2$, see Eq. \eqref{eq_def_fraction}, which in the LLL projection leads to ideal GKP states in the absence of the parabolic confinement potential. As in the previous section, we construct and analyze a low energy theory of the system, valid in the weak Landau level coupling limit.

\subsection{Numerical Results}
\begin{figure}[t]
	\centering
	\includegraphics[width=0.5\textwidth]{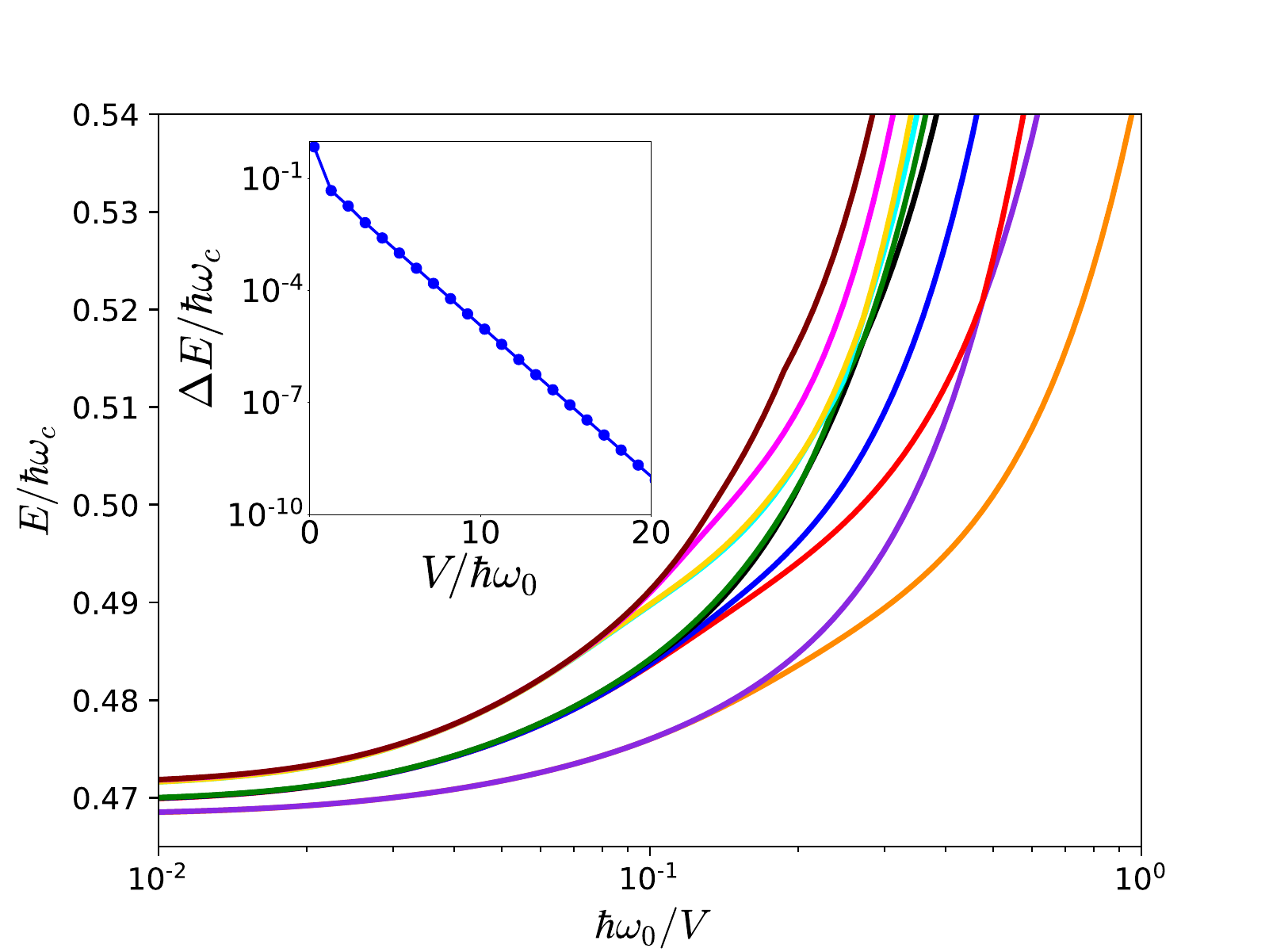}
	\caption{Lowest ten eigenenergies of the Hamiltonian in Eq. \eqref{eq_confined_Hamiltonian_initial} as functions of the confinement strength $\hbar\omega_0/V$ for $p/q=1/2$ and $V/\hbar\omega_c=0.25$. Inset: energy gap $\Delta E = E_1 - E_0$ between the two lowest eigenstates for the same values of $p/q$ and $V/\hbar \omega_c$. For $V/\hbar\omega_0 \gg 1$, the energy gap decreases exponentially with $ V/\hbar \omega_0$. }
	\label{fig_lowest_10_energies}
\end{figure}

When a confinement potential is included, the two-fold degeneracy of the ground space is lifted and an energy gap opens between the degenerate ground states of $H_\text{LLL}$ in Eq. \eqref{eq_eff_GKP_Hamiltonian}. This energy gap, however, is exponentially small in $\omega_0$, and so the ground state and the first excited state remain quasi-degenerate when $\omega_0$ is small enough.
To illustrate this point, in Fig. \ref{fig_lowest_10_energies}, we show the lowest ten eigenenergies of the Hamiltonian in Eq. \eqref{eq_confined_Hamiltonian_initial} as functions of the confinement strength $\hbar\omega_0/V$ for a rather large value of $V/\hbar\omega_c=0.25$. Recall that $V$ denotes the amplitude of $V_\text{crys}$ in Eq. \eqref{eq_V_crys}.
We observe that the energy gap between the ground state and first excited state is negligibly small up to confinements of $\hbar\omega_0 / V \lesssim 0.2$, and that it nicely fits an exponential scaling $E_1-E_0 \propto \exp(-\alpha V / \hbar \omega_0)$, with a positive constant prefactor $\alpha$.
In addition, the spectrum is now discrete and higher excited states are gapped from the two-fold quasi-degenerate ground space. 

As mentioned in Sec. \ref{sec_GKP}, the discreteness of the spectrum then allows one to use a perturbative argument which states that local perturbations do not considerably alter the spectrum of the Hamiltonian.

\begin{figure}[h!]
	\centering
	\includegraphics[width=0.5\textwidth]{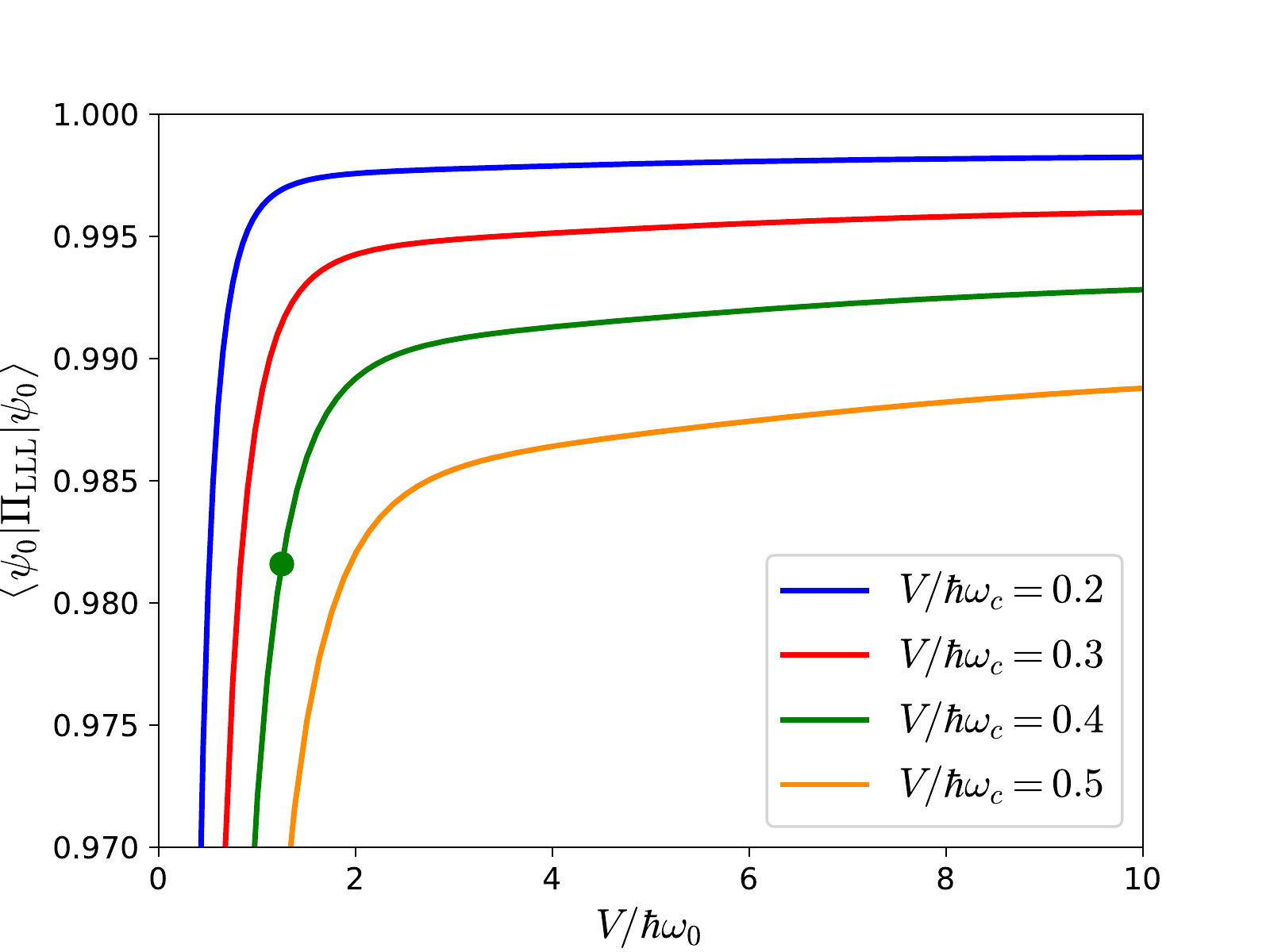}
	\caption{Expectation value of the LLL projector in the ground state of the Hamiltonian in Eq. \eqref{eq_confined_Hamiltonian_initial} for different values of $V/\hbar\omega_c$ as function of $V/\hbar\omega_0$. As expected, the LLL projection becomes more accurate for smaller values of the energy ratios $V/\hbar\omega_c$ and $\hbar\omega_0/V$. We mark with a green circle the values of $V/\hbar\omega_c=0.4$ and $\hbar\omega_0/V=0.8$ that are experimentally relevant for our circuit proposal, see Sec. \ref{sec_circuit}. For these values, we find $\bra{\psi_0}\Pi_\text{LLL} \ket{\psi_0} = 0.981$.}
	\label{fig_validity_LLL_projection}
\end{figure}

\begin{figure}[t]
	\centering
	\includegraphics[width=0.48\textwidth]{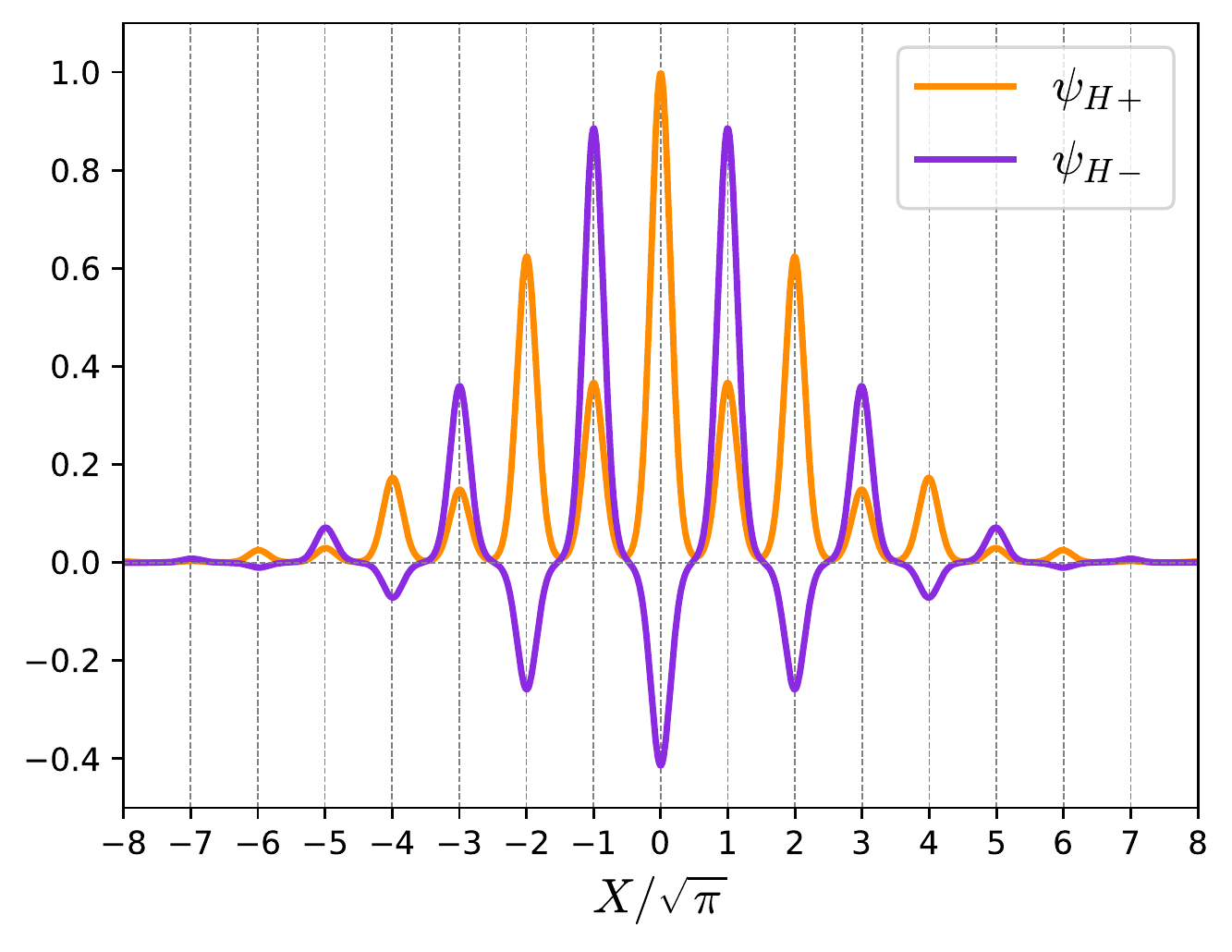}
	\caption{Lowest-energy eigenfunctions of the effective Hamiltonian in Eq. \eqref{eq_H_parabolic_projected_with_x_p} obtained by using $\hbar \omega_0^2 / \omega_c  V_0= 0.05$, which corresponds to $\Delta=0.25$, see Eq. \eqref{eq_width_delta}. The quasi-degenerate eigenfunctions are even and odd under Fourier transforming, respectively, and are well approximated by Eq. \eqref{eq_lin_combination}.}
	\label{fig_almost_grid_states}
\end{figure}

Because we are interested in the weak Landau level coupling limit, we analyze the effect of higher Landau levels on the low energy eigenstates numerically. In Fig. \ref{fig_validity_LLL_projection} we show how the expectation value of the LLL projector $\Pi_\text{LLL} = \ketbra{0}_{\bm{\pi}}$ in the ground state of the Hamiltonian in Eq. \eqref{eq_confined_Hamiltonian_initial} varies as a function of the confinement potential for fixed values of $V/\hbar\omega_c$. We observe that the higher Landau levels have a negligible effect for a wide range of parameters, giving an error below $3\%$ for rather large values of both $V/\hbar\omega_c$ and $\hbar\omega_0/V$, and, consequently, justifying even in this case a projection onto the LLL. In particular, at the values $V/\hbar\omega_c=0.4$ and $\hbar\omega_0/V = 0.8$, which are the relevant parameters for the circuit model presented in Sec. \ref{sec_circuit}, we find $\bra{\psi_0}\Pi_\text{LLL} \ket{\psi_0} = 0.981$, see the green circle in Fig. \ref{fig_validity_LLL_projection}.

The effective Hamiltonian of the system in this limit is analyzed in the next section.

\subsection{Approximate Grid States in the LLL Projection}\label{subsec_low_energy}
In analogy to Sec. \ref{subec_LLL_projection}, here we find an effective Hamiltonian that captures the behavior of the system in the weak Landau level coupling limit. By projecting Eq. \eqref{eq_confined_Hamiltonian_initial} onto the LLL and considering $p/q=1/2$,  we obtain 
\begin{equation}\label{eq_H_parabolic_projected_with_x_p}
	H_\text{LLL} 
	= \frac{\hbar \omega_0^2}{\omega_c} \frac{P^2 + X^2}{2}
	- V_0 \bigg[ \cos(2 \sqrt{\pi} X) + \cos(2 \sqrt{\pi} P) \bigg],
\end{equation}
where $V_0 = V e^{-\pi}$ and the canonical position $X$ and momentum $P$ are defined by Eq. \eqref{eq_X_P_rescaling}.

\begin{figure}[t!]
	\centering
	\includegraphics[width=0.48\textwidth]{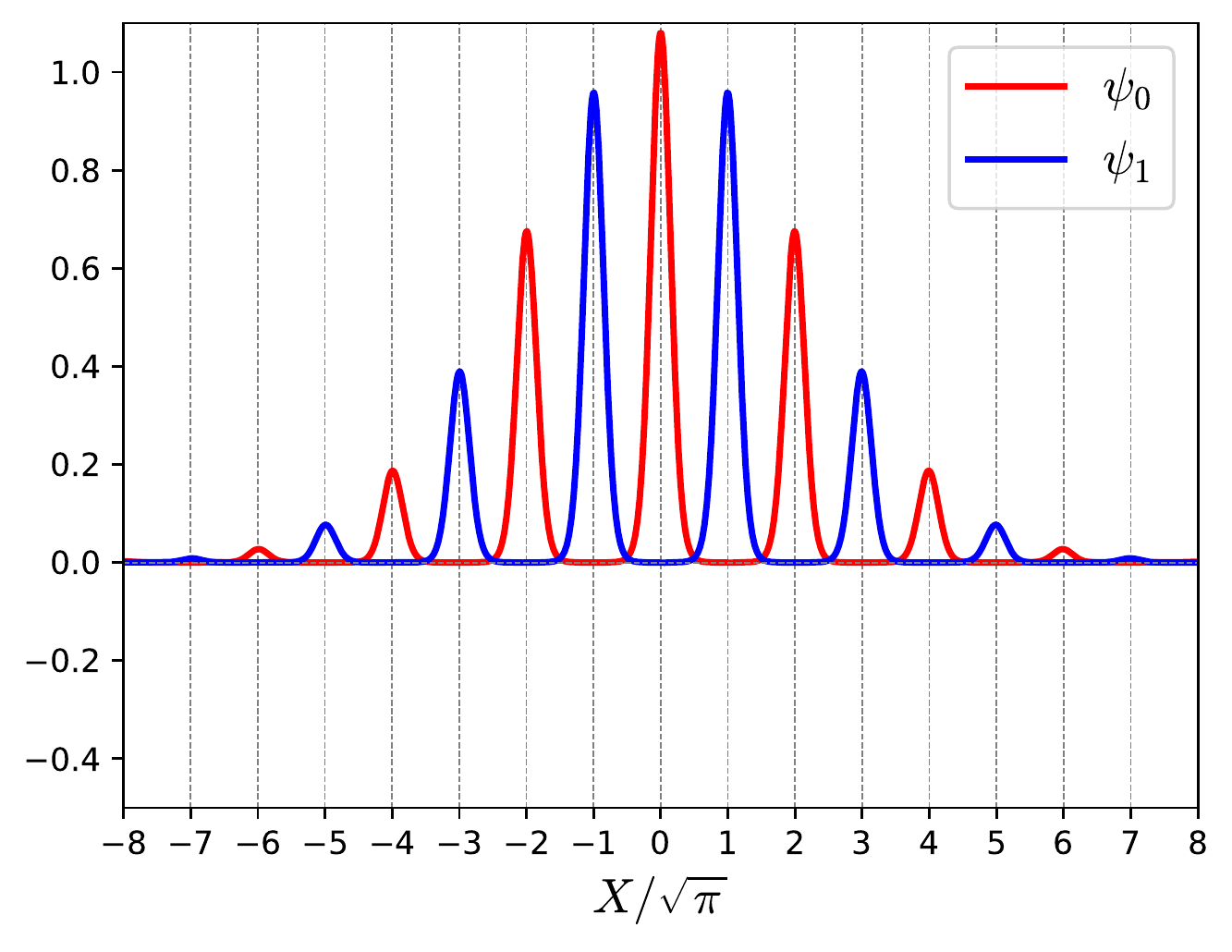}
	\caption{The approximate grid states in Eq. \eqref{eq_fit_approx_grid_states} obtained using the same parameters as in Fig. \ref{fig_almost_grid_states}, $\hbar \omega_0^2 / \omega_c V_0  = 0.05$, which corresponds to $\Delta=0.25$. The widths of the individual peaks and of the total envelope are the inverse of each other.}
	\label{fig_approx_grid_states}
\end{figure}

Importantly, the parabolic potential in Eq. \eqref{eq_total_potential} reduces to the harmonic oscillator Hamiltonian with frequency $\omega_0^2/\omega_c$ after the LLL projection, and breaks the periodicity of both $X$ and $P$. 
However, note that the confinement potential preserves the four-fold rotational symmetry in $x_1$ and $x_2$. Because a $\pi/2$ rotation in the $x_1x_2$-plane reduces to a Fourier transform, which maps $X \mapsto P$ and $P \mapsto -X$, when projected onto the LLL, the Hamiltonian in Eq. \eqref{eq_H_parabolic_projected_with_x_p} is invariant under the exchange of $X$ and $P$ and its eigenstates are also eigenstates of the Fourier transform. More detailed explanations of this symmetry and correspondence are given in Appendices \ref{app_numerical_preparation} and \ref{app_4fold_rotation}.

In particular, the two quasi-degenerate low-energy eigenfunctions $\psi_{H+}(X)$ (ground state) and $\psi_{H-}(X)$ (first excited state), shown in Fig. \ref{fig_almost_grid_states}, are even and odd eigenfunctions of the Fourier transform with eigenvalues $\pm 1$, respectively. These states are well approximated by the linear combinations
\begin{subequations}\label{eq_lin_combination}
\begin{align}
	\psi_{H+}(X) &\approx \cos(\frac{\pi}{8}) \psi_0(X) + \sin(\frac{\pi}{8}) \psi_1(X), 		\\
	\psi_{H-}(X) &\approx -\sin(\frac{\pi}{8}) \psi_0(X) + \cos(\frac{\pi}{8}) \psi_1(X),
\end{align}
\end{subequations}
of the approximate grid states
\begin{subequations}\label{eq_fit_approx_grid_states}
\begin{align}
	\psi_0(X) &= \frac{\sqrt{2}}{\pi^{1/4}} e^{- X^2 \Delta^2 /2}
	\sum_{n=-\infty}^\infty \exp(- \tfrac{(X-2\sqrt{\pi}n)^2}{2 \Delta^2}),			\\
	\psi_1(X) &= \frac{\sqrt{2}}{\pi^{1/4}} e^{- X^2 \Delta^2 /2}
	\sum_{n=-\infty}^\infty \exp(- \tfrac{(X-2\sqrt{\pi}n -\sqrt{\pi})^2}{2 \Delta^2}),
\end{align}
\end{subequations}
that are shown in Fig. \ref{fig_approx_grid_states}.  Explicitly, the squeezing parameter $\Delta$ is given by
\begin{equation}\label{eq_width_delta}
	\Delta = \left( \frac{\hbar \omega_0^2}{4 \pi \omega_c V_0} \right)^{1/4} \ll 1.
\end{equation}
These approximate grid states are obtained by a convolution of the ideal grid states in Eq. \eqref{eq_GKP_code_words} with a narrow Gaussian of width $\Delta$ and a multiplication with a wide Gaussian of width $1/\Delta$. This inverse relation is a consequence of the invariance of the Hamiltonian with respect to the Fourier transform and it also reflects the fact that the GKP code corrects equal errors in the $X$- and $P$-variable. 

When the confinement frequency $\omega_0$ is decreased, the width $\Delta$ of the individual Gaussian peaks of the approximate grid states in Eq. \eqref{eq_fit_approx_grid_states} decreases, while the broadening of the envelope function increases, eventually recovering the ideal grid states in Eq. \eqref{eq_GKP_code_words} when $\omega_0 \rightarrow 0$.

The derivation of Eqs. \eqref{eq_lin_combination} - \eqref{eq_width_delta} is based on a nested application of the envelope function approximation \cite{Girvin, Kittel, Rossi} discussed in Appendix \ref{appendix_envelope_function_theory}. 
Note that, when the parameter $\Delta$ is rather small, as we are considering here, the states $\psi_0(X)$ and $\psi_1(X)$ are orthonormal up to an exponentially small correction scaling as $\sim e^{-1/\Delta^2}$. Consequently, they form an appropriate computational basis for the quasi-degenerate ground space of $H_\text{LLL}$.
 
The angle $\pi/8$ which appears in the linear combination in Eq. \eqref{eq_lin_combination} can be understood by considering that in the basis $\psi_{0,1}(X)$, the Fourier transform approximately equals the Hadamard gate \cite{GKP}
\begin{equation}\label{eq_H_gate}
	\overline{H}
	= \frac{1}{\sqrt{2}}
	\begin{pmatrix}
		1	&	1	\\
		1	&	-1
	\end{pmatrix},
\end{equation}
whose even and odd eigenfunctions are $\psi_{H+}(X)$ and $\psi_{H-}(X)$, respectively \footnote{On the Bloch sphere, the Hadamard gate corresponds to a $\pi$-rotation around the axis defined by $\overline{X}+\overline{Z}$.}. These states are magic states which combined with Clifford operations achieve universal quantum computation \cite{GKP, BravyiKitaev, Yamasaki, Menicucci}. \\

We remark that an alternative low-energy description of the Hamiltonian in Eq. \eqref{eq_confined_Hamiltonian_initial} which relies on the introduction of the eigenbasis of the quadratic part of the Hamiltonian is possible. This basis is known as Fock-Darwin basis \cite{Fock, Darwin}. In the weak Landau level coupling limit, the results obtained with this approach are equivalent to the ones shown here.

\section{Eigenfunctions of the two-dimensional problem}\label{sec_full_wf}
So far, we described eigenfunctions of an one-dimensional Hamiltonian obtained by projecting a two-dimensional Hamiltonian onto the LLL. Here, we establish the connection between the eigenfunctions of these two Hamiltonians. In the weak Landau level coupling regime, which we considered in the previous sections, the wave function of the two-dimensional system is the coherent state representation of the wave function of the one-dimensional system.
In the following, we begin by considering the ideal case discussed in Sec. \ref{sec_crystal} and then we straightforwardly generalize our result to include the parabolic confinement potential as introduced in Sec. \ref{sec_confinement}. \\

In the weak Landau level coupling limit, the low-energy eigenstates of the Hamiltonian in Eq. \eqref{eq_final_crystal_Hamiltonian} are well approximated by the product state
\begin{equation}\label{eq_tensor_2d_wavefunction}
	\ket{\Psi} = \ket{0}_\pi \otimes \ket{\psi}_R \equiv \ket{0,\psi},
\end{equation}
where $\ket{0}_\pi$ denotes the LLL and $\ket{\psi}_R$ is an eigenstate of the projected Hamiltonian in Eq. \eqref{eq_eff_GKP_Hamiltonian}. From Eq. \eqref{eq_tensor_2d_wavefunction}, it follows that the wave function $\Psi(x_1, x_2) = \braket{x_1, x_2}{\Psi}$ in the original coordinates $x_i$ and the one-dimensional wave function $\psi(X) = \braket{X}{\psi}_R$ are related by the unitary integral transform
\begin{equation}\label{eq_integral_transform}
	\Psi(x_1, x_2)	= \int_{-\infty}^\infty \! dX K_0(x_1, x_2 ; X) \psi(X),
\end{equation}
where $K_0(x_1, x_2 ; X) = \braket{x_1, x_2}{0, X}$ is a gauge-dependent integration kernel. As derived in Appendix \ref{app_kernel}, in the symmetric gauge, i.e. $\bm{A}(x_1, x_2)=B/2(-x_2, x_1, 0)^T$, we obtain
\begin{equation}\label{eq_integration_kernel}
\begin{split}
	K_0(x_1, x_2 ; X)
	= \frac{1}{\sqrt{2}\pi^{3/4}}  \exp( - \frac{(X - x_1)^2}{2} ) \\ 
	\times \exp(-i x_2 X) \exp( i \frac{x_1 x_2}{2} ).
\end{split}
\end{equation}
To simplify the notation, in this section, we work in magnetic units and we rescale all the lengths by the magnetic length $l_B= \sqrt{\hbar/eB}$.

Note that, up to a gauge phase $\exp(ix_1x_2/2)$, the integration kernel $K_0(x_1, x_2 ; X)$ is the complex conjugate of the wave function (in $X$-representation) of a coherent state with average position and momentum $x_1$ and $x_2$, respectively. Consequently, as long as the matrix elements between different Landau levels are small and the approximate factorization in Eq. \eqref{eq_tensor_2d_wavefunction} is valid, the low-energy two-dimensional eigenfunctions of the Hamiltonian in Eq. \eqref{eq_final_crystal_Hamiltonian} are the coherent state representations of the eigenfunctions of the projected Hamiltonian in Eq. \eqref{eq_eff_GKP_Hamiltonian} \cite{Dariusz}.
It follows that the absolute value squared $|\Psi(x_1, x_2)|^2$ is the non-negative Husimi $Q$ representation \cite{Walls, Gerry, Subramanyan} associated with $\psi(X)$.
Note also that the integral transform in Eq. \eqref{eq_integral_transform} is invertible and preserves orthonormality.

Let us consider now the LLL non-normalizable eigenfunction
\begin{equation}\label{eq_crystal_Zak_states}
	\psi_{\bm{k}}(X) = e^{-i k_1  X} \sum_{n \in \mathbb{Z}} \delta(X - 2\sqrt{\pi}n - k_2),
\end{equation}
of the effective Hamiltonian in Eq. \eqref{eq_eff_GKP_Hamiltonian}, satisfying the quasi-periodic boundary conditions defined by Eq. \eqref{eq_toroidal_2} with $p/q=1/2$ and $l=0$. Because of the choice $p/q=1/2$, the length $L_0$ (in magnetic units) reduces to $\sqrt{\pi}$, see Eq. \eqref{eq_def_fraction}. The wave function of the two-dimensional system, obtained via the integral transform in Eq. \eqref{eq_integral_transform}, is
\begin{equation}\label{eq_Haldane_Rezayi}
\begin{split}
	\Psi_{\bm{k}}(x_1, x_2) = &\frac{1}{\sqrt{2}\pi^{3/4}}
	e^{-\frac{x_1(x_1-ix_2)}{2}}																						\\
	&\times \vartheta \begin{bmatrix} k_2/2\sqrt{\pi} \\ -k_1/\sqrt{\pi} \end{bmatrix} 
	\left( \frac{-i(x_1-ix_2)}{\sqrt{\pi}}, 2i \right),
\end{split}
\end{equation}
with the generalized elliptic theta function
\begin{equation}\label{eq_elliptic_theta}
	\vartheta \begin{bmatrix} a \\ b \end{bmatrix} (z, \tau)
	= \sum_{n \in \mathbb{Z}} e^{i \pi (n+a)^2\tau} e^{i2\pi(n+a)(z+b)}.
\end{equation}
Note that the absolute value of the wave function in Eq. \eqref{eq_Haldane_Rezayi} is periodic in both $x_1$ and $x_2$, i.e. $|\Psi_{\bm{k}}(x_1, x_2)| = |\Psi_{\bm{k}}(x_1+2\sqrt{\pi}, x_2)| = |\Psi_{\bm{k}}(x_1, x_2+\sqrt{\pi})|$. Similar two-dimensional functions satisfying quasi-periodic boundary conditions were introduced by Haldane and Rezayi \cite{Rezayi} as a basis to describe the problem of an electron confined to the surface of a torus and under the effect of a perpendicular magnetic field. 

\begin{figure}[t]
	\centering
	\subfloat[]{\includegraphics[width=0.50\textwidth]{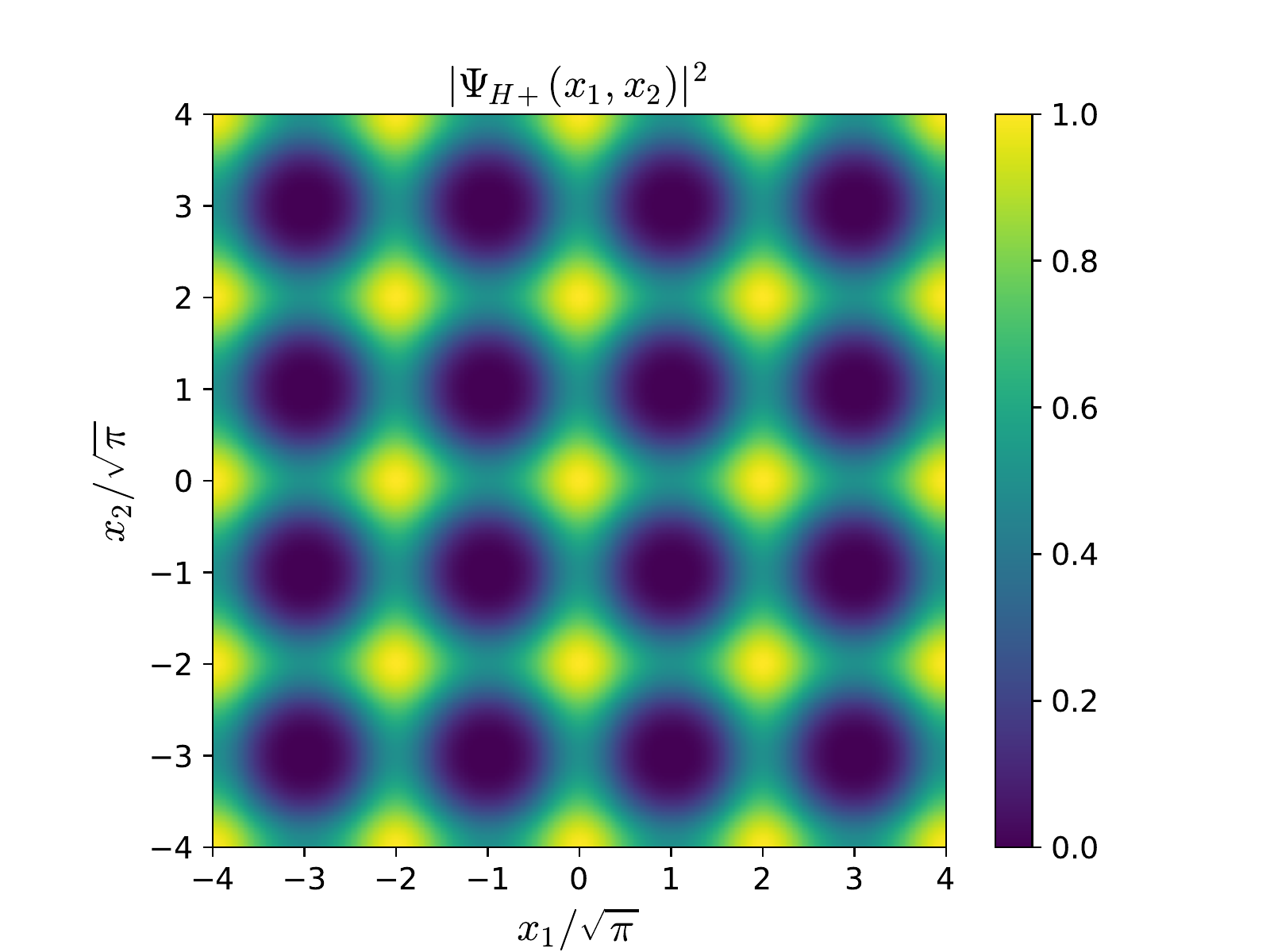}}\\
	\subfloat[]{\includegraphics[width=0.50\textwidth]{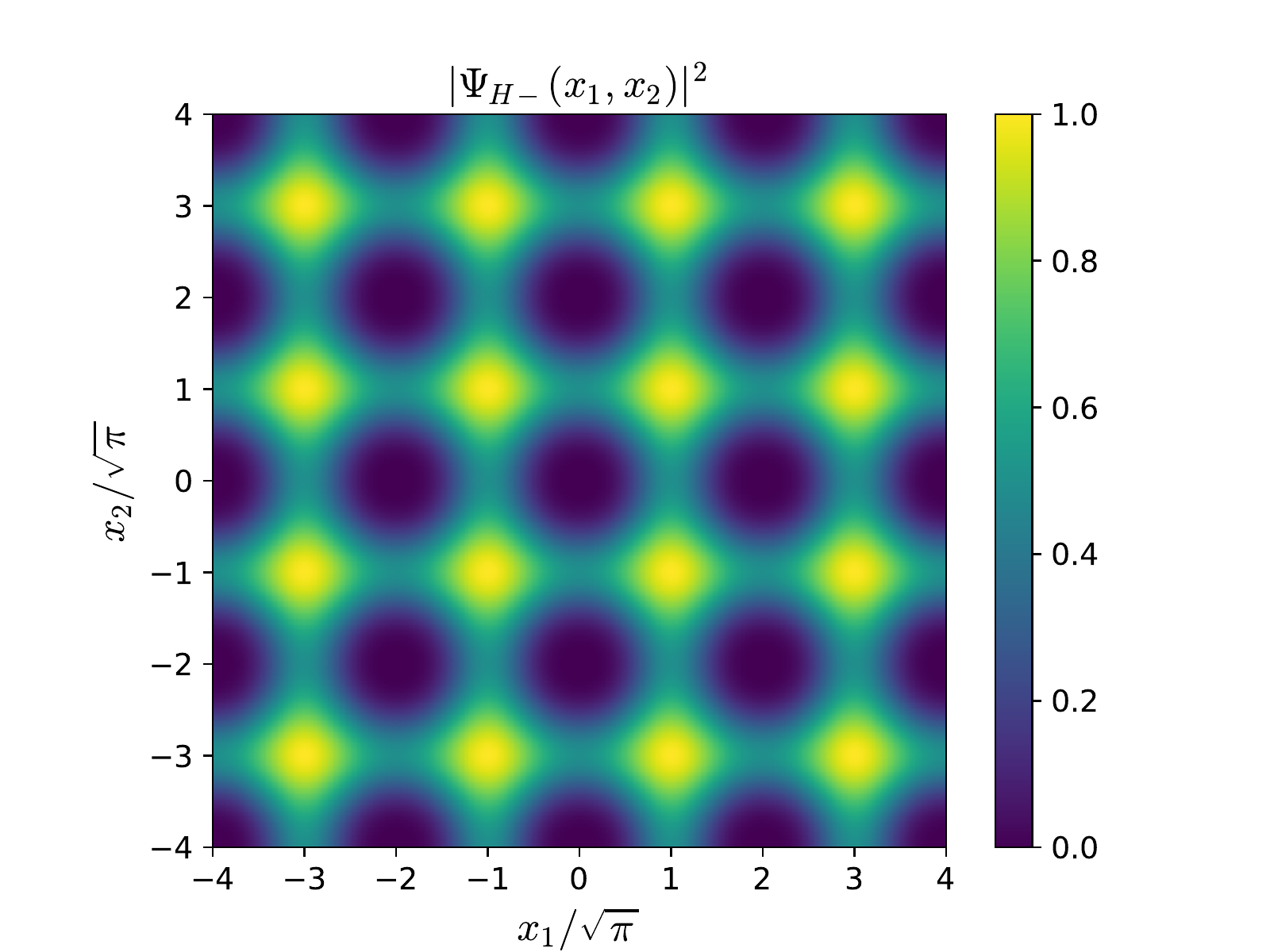}}	
\caption{Absolute values squared of the wave functions a) $\Psi_{H+}(x_1,x_2)$ and b) $\Psi_{H-}(x_1,x_2)$. These wave functions approximate the low energy eigenstates of the two-dimensional Hamiltonian in Eq. \eqref{eq_final_crystal_Hamiltonian} and are constructed to be even and odd under a $\pi/2$-rotation in the $x_1x_2$-plane. Note that the wave functions in the plot are not normalized. We remark that $|\Psi_{H\pm}(x_1, x_2)|^2$ are the Husimi Q quasi-probability functions associated with the eigenstates $\psi_{H\pm}(X)$ [defined in Eq. \eqref{eq_ideal_2dim_wf}] of the projected Hamiltonian in Eq. \eqref{eq_eff_GKP_Hamiltonian}.}
	\label{fig_magic_full_wf}
\end{figure}

The two-fold degenerate ground space of the effective Hamiltonian in Eq. \eqref{eq_eff_GKP_Hamiltonian} is spanned by the logical codewords in Eq. \eqref{eq_GKP_code_words}, which are obtained from Eq. \eqref{eq_crystal_Zak_states} by considering $\bm{k}=(0,0)^T$ and $\bm{k}=(0, \sqrt{\pi})^T$, respectively. As discussed in Sec. \ref{subsec_low_energy} and in Appendix \ref{app_4fold_rotation}, $\pi/2$-rotations in the $x_1x_2$-plane correspond to a Fourier transform after the LLL projection.
Consequently, because in the basis $\psi_0(X)$ and $\psi_1(X)$ a Fourier transform is equivalent to a Hadamard gate \cite{GKP} (see Sec. \ref{subsec_low_energy}), to construct four-fold rotational symmetric wave functions in the two-dimensional plane, we consider the linear combinations \footnote{Note that Eq. \eqref{eq_ideal_2dim_wf} is exact because the ideal GKP states $\psi_{0,1}(X)$ are considered. In contrast, in Sec. \ref{subsec_low_energy} we considered approximate grid states.}
\begin{subequations}\label{eq_ideal_2dim_wf}
\begin{align}
	\psi_{H+}(X) &= \cos(\frac{\pi}{8}) \psi_0(X) + \sin(\frac{\pi}{8}) \psi_1(X),\\
	\psi_{H-}(X) &= -\sin(\frac{\pi}{8}) \psi_0(X) + \cos(\frac{\pi}{8}) \psi_1(X).
\end{align}
\end{subequations}
These functions are even (odd) under Fourier transform and so the corresponding two-dimensional wave functions $\Psi_{H\pm}(x_1,x_2)$ are even (odd) under a $\pi /2$-rotation around the origin $x_i=0$. 
The absolute values of the functions $\Psi_{H\pm}(x_1,x_2)$ are shown in Fig. \ref{fig_magic_full_wf}. We observe that the absolute values are periodic with period $2\sqrt{\pi}$ in both $x_1$- and $x_2$-direction. Also, we find that these states are related to each other by
\begin{equation}
	\Psi_{H+}(x_1,x_2) = \Psi_{H-}(x_1+\sqrt{\pi},x_2+\sqrt{\pi}) e^{- i \sqrt{\pi} \frac{1+(x_1-x_2)}{2}},
\end{equation}
and so the absolute values of the two wave functions are simply obtained by a shift of $\sqrt{\pi}$ in the $x_1$- and $x_2$-direction. \\

\begin{figure}[t]
	\centering
	\subfloat[]{\includegraphics[width=0.50\textwidth]{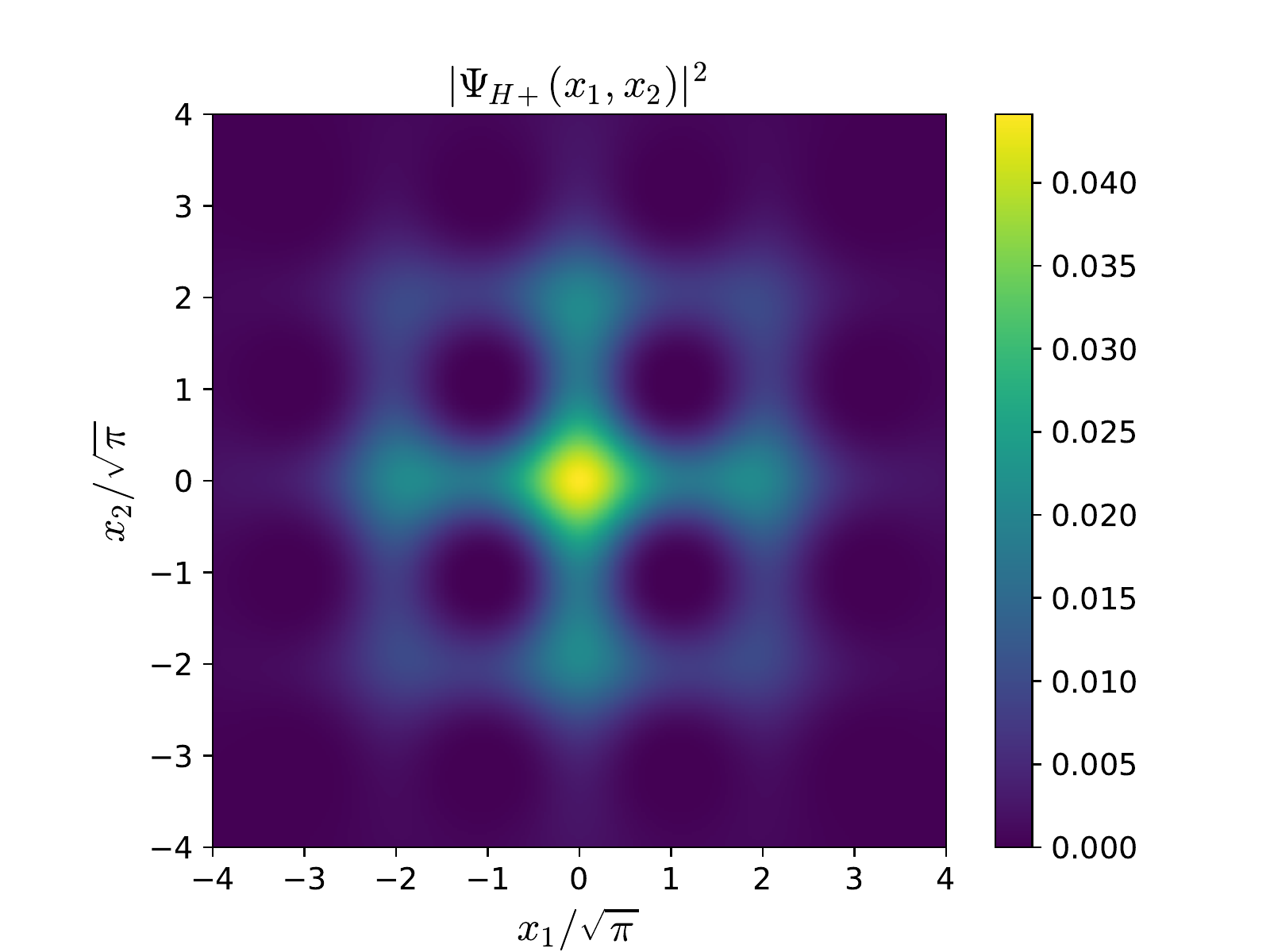}}\\
	\subfloat[]{\includegraphics[width=0.50\textwidth]{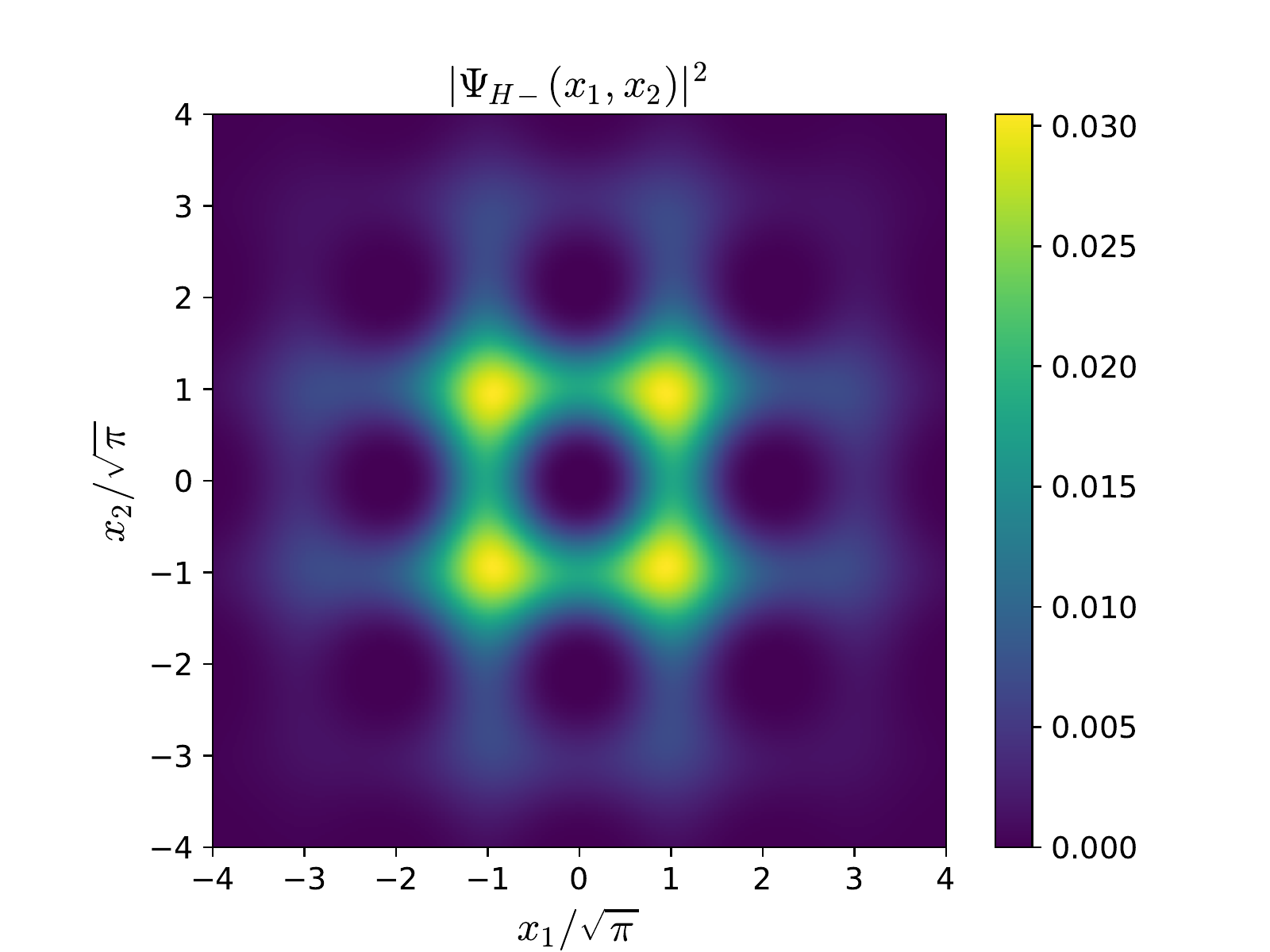}}
	\caption{Absolute values squared of the quasi-degenerate ground state wave functions a) $\Psi_{H+}(x_1, x_2)$ and b) $\Psi_{H-}(x_1, x_2)$  of the Hamiltonian in Eq. \eqref{eq_confined_Hamiltonian_initial}, which includes a parabolic confinement potential. These functions are obtained by combining Eqs. \eqref{eq_lin_combination} and \eqref{eq_psi01_theta} and are even and odd under a $\pi/2$-rotation in the $x_1x_2$-plane. The wave functions here are normalized and are obtained by using $\Delta=0.25$.}
	\label{fig_magic_full_wf_conf}
\end{figure}

As long as the Landau level coupling remains weak, Eq. \eqref{eq_integral_transform} is appropriate to describe also the system discussed in Sec. \ref{sec_confinement}, where an additional parabolic potential is included. In particular, we find that the approximate grid states given in Eq. \eqref{eq_fit_approx_grid_states} transform into
\begin{equation}\label{eq_psi01_theta}
\begin{split}
	\Psi_j(x_1, x_2) = &\sqrt{\tfrac{2\Delta^2}{\pi(1+\Delta^2+\Delta^4)}}
	e^{\frac{\Delta^2(x_1-ix_2)^2}{2(1+\Delta^2+\Delta^4)}} e^{-\frac{x_1(x_1-ix_2)}{2}}		\\
	&\times \vartheta \begin{bmatrix} \frac{j}{2} \\ 0 \end{bmatrix} 
	\left(
	\tfrac{-i(x_1-ix_2)}{\sqrt{\pi}(1+\Delta^2+\Delta^4)}, 2i \tfrac{1+\Delta^2}{1+\Delta^2+\Delta^4}
	\right),
\end{split}
\end{equation}
with $j=0,1$ and $\Delta$ being defined in Eq. \eqref{eq_width_delta}. The low-energy eigenstates $\Psi_{H\pm}(x_1, x_2)$ of the Hamiltonian in Eq. \eqref{eq_confined_Hamiltonian_initial} are related to these basis states by Eq. \eqref{eq_lin_combination}, and their absolute values obtained for $\Delta=0.25$ are shown in Fig. \ref{fig_magic_full_wf_conf}.

\newpage
Comparing Figs. \ref{fig_magic_full_wf} and \ref{fig_magic_full_wf_conf}, we observe that the parabolic potential introduces a Gaussian decay roughly of the order $1/\Delta^2$ of the wave functions in both $x_1$- and $x_2$-direction and also it distorts the arguments of the theta functions with corrections of order $\Delta^2$.
Of course, the absolute values of the wave functions in Fig. \ref{fig_magic_full_wf} are recovered by taking the limit $\Delta \rightarrow 0$.

\newpage
\section{GKP Hamiltonian in a non-reciprocal superconducting circuit}\label{sec_circuit}
\begin{figure}[t]
	\centering
	\includegraphics[width=0.48\textwidth]{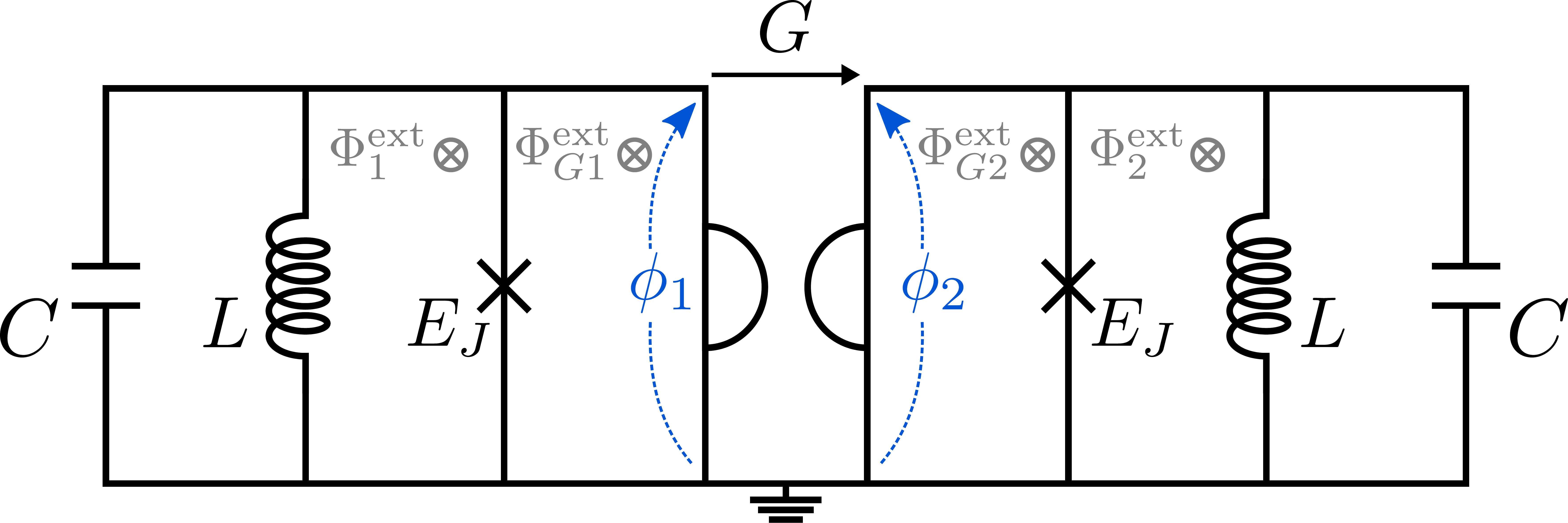}
	\caption{Circuit design implementing the Hamiltonian in Eq. \eqref{eq_LCJJ_G_LCJJ_Hamiltonian}, which approximates the GKP Hamiltonian.}
	\label{fig_circuit}
\end{figure}

Here, we propose a possible experimental realization of the Hamiltonian in Eq. \eqref{eq_confined_Hamiltonian_initial} based on a combination of state-of-the-art non-reciprocal superconducting circuits. We consider here the circuit shown in Fig. \ref{fig_circuit}.

The device consists of two fluxonia  coupled by a gyrator.
The fluxonium is a well-known superconducting circuit comprising a Josephson junction with Josephson energy $E_J$ in parallel with a capacitance $C$ and an inductance $L$ \cite{Manucharyan, ManucharyanX} \footnote{Strictly speaking, this superconducting circuit is only called to be a fluxonium if it is operated in the parameter regime $E_L \ll E_J$ and $1 \lesssim E_J/E_C \lesssim 10$. However, regardless of this choice of parameters and for the sake of convenience, we refer to the circuit as fluxonium.}. The crucial difference of our design from more conventional superconducting qubit architectures is the non-reciprocity that comes from the gyrator \cite{Tellegen}.

A gyrator is a two-port linear device that relates incoming currents and voltages according to 
\begin{equation}\label{eq_gyrator_admittance}
	\begin{pmatrix}
	I_1		\\	I_2
	\end{pmatrix}
	=
	\underbrace{	\begin{pmatrix}
	0	&	-G \\
	G	&	0
	\end{pmatrix}}_{\bm{Y}}
	\begin{pmatrix}
	V_1		\\	V_2
	\end{pmatrix},
\end{equation}
where $G$ is the frequency-independent gyration conductance. Because it is characterized by an anti-symmetric admittance matrix $\bm{Y}$, this device is non-reciprocal and breaks the time-reversal symmetry of the circuit.

While the typical implementations of these devices are quite bulky \cite{Hogan1, Hogan2}, there are also recent realizations of miniaturized on-chip non-reciprocal devices based on actively pumped systems \cite{Rosenthal, Chapman, Lecocq, barzanjeh2017} or based on the quantum (anomalous) Hall effect \cite{Reilly, Mahoney}. Although our model is independent of the specific realization of the gyrator, the latter devices are advantageous in this context because they are passive and they rely on quantized excitations with a long lifetime that can be well-described by the theory of circuit quantum electrodynamics (cQED) \cite{Vool, Girvin_cQED}. A further advantage will be the value of $G$.

To describe the system with circuit quantization theory, we introduce the node fluxes $\phi_i (t) = \int_{-\infty}^{t} V_i(t') dt'$. Using Kirchoff's laws, the following contribution to the Lagrangian \cite{Rymarz, Duinker}
\begin{equation}\label{eq_Lagrangian_gyrator_symmetric_gauge}
	\mathcal{L}_G = \frac{G}{2} \left( \phi_1 \dot{\phi}_2 - \dot{\phi}_1 \phi_2 \right),
\end{equation}
correctly reproduces the defining property of the gyrator in Eq. \eqref{eq_gyrator_admittance} when two general classical networks are attached to it. Importantly, note that (in the style of an electronic system) Eq. \eqref{eq_Lagrangian_gyrator_symmetric_gauge} is similar to the effect of a homogeneous magnetic field of strength $B=G/e$ passing through the $\phi_1\phi_2$-plane. 
More details on circuit quantization of non-reciprocal devices can be found in Refs. \cite{Rymarz, Parra}.

For now, we neglect the effect of magnetic fluxes threading the superconducting loops and we set $\Phi^\text{ext}_i = \Phi^\text{ext}_{Gi} = 0$. Combining conventional circuit QED with Eq. \eqref{eq_Lagrangian_gyrator_symmetric_gauge}, we find the Hamiltonian of the circuit in Fig. \ref{fig_circuit} to be 
\begin{equation}\label{eq_LCJJ_G_LCJJ_Hamiltonian}
\begin{split}
	H = &\frac{(Q_1+G\phi_2/2)^2}{2C} + \frac{(Q_2-G\phi_1/2)^2}{2C} 
		+ \frac{1}{2L} (\phi_1^2 + \phi_2^2) 						\\
	&- E_J \bigg[ \cos \left(\frac{2\pi}{\Phi_{0,s}} \phi_1 \right)
				\hspace{0pt} + \cos \left(\frac{2\pi}{\Phi_{0,s}} \phi_2 \right) \bigg].
\end{split}
\end{equation}
We then impose the canonical commutation relation $[\phi_i, Q_j] = i \hbar \delta_{ij}$. Here, $Q_i$ are the charges on the $i$'th capacitor. Note that the superconducting flux quantum $\Phi_{0,s} =h/2e$ differs from the flux quantum $\Phi_0$ used in the previous sections by a factor $2$. For simplicity, we assumed here that the two fluxonia coupled to the gyrator are identical. We do not expect small anisotropies to drastically alter the results described in this section.

The Hamiltonian in Eq. \eqref{eq_LCJJ_G_LCJJ_Hamiltonian} describing our circuit has the same structure as the Hamiltonian of a confined crystal electron in a magnetic field in Eq. \eqref{eq_confined_Hamiltonian_initial} and discussed in detail in Sec. \ref{sec_confinement}. The variables that play equivalent roles in the two cases are given in Table \ref{tab_identification}.

\setlength{\tabcolsep}{4 pt}
\begin{table}[h]
   \centering
	\begin{tabular}{ c || c | c | c | c | c | c| c }
	\textbf{Crystal electron} &  $x_i$  &  $p_i$  &  $m$  
	&  $eB$  &  $V$  &  $L_0$  &  $\omega_0$ \\   
	\hline 
	\textbf{Circuit} &  $\phi_i$  &  $Q_i$  &  $C$  
	&  $G$  &  $E_J$  &  $\Phi_{0,s}$  &  $\omega_{LC}$   
	\end{tabular} 
   \caption{Mapping of the parameters and variables used in the jargon of a crystal electron and cQED, such that the Hamiltonians in Eqs. \eqref{eq_confined_Hamiltonian_initial} and \eqref{eq_LCJJ_G_LCJJ_Hamiltonian} coincide. Note that the cyclotron frequency in terms of circuit parameters is $\omega_c = G/C$.}
   \label{tab_identification}
\end{table}

In particular, the gyration conductance $G$ acts as a magnetic field $B$ and the  characteristic frequency of the LC circuit $\omega_{LC}=1/\sqrt{L C}$ acts as the harmonic confinement $\omega_0$. For later convenience, we also introduce the charging energy $E_C = e^2/2C$ and the inductive energy $E_L = \Phi_{0, s}^2 / 4 \pi^2 L$. 

As shown in Sec. \ref{sec_crystal}, the number of flux quanta threading one unit cell is of fundamental importance for realizing GKP states. In our circuit, Eq. \eqref{eq_def_fraction} becomes
\begin{equation}
	\frac{p}{q} = \frac{G}{e} \frac{\Phi_{0,s}^2}{\Phi_0} = \frac{G}{G_0},
\end{equation}
where we introduced the superconducting conductance quantum $G_0 = (2e)^2/h$. In order to obtain GKP states, we require $p/q=1/2$ and, accordingly, we require the gyration conductance to be precisely
\begin{equation}\label{eq_desired_G}
	G = \frac{2e^2}{h}.
\end{equation}
We remark again that while this value of $G \sim 1/(13 \, \text{k}\Omega)$ is unrealistic for superconducting based gyrators, it can be easily reached using quantum (anomalous) Hall effect devices, where the characteristic impedance is $1/G=h/2e^2\nu$ \cite{Viola, Bosco1, Bosco2, BoscoReilly, Mahoney, Bestwick, Reilly}, with $\nu$ being the Landau level filling factor. The robust quantization of the Hall conductivity in these materials also guarantees that the value of $G$ remains precisely fixed for a wide range of design parameters, hence improving the reproducibility of the gyrator.

To reach low values of the harmonic confinement frequency $\omega_{LC}$, we expect that the novel hyperinductances \cite{Ray}, the kinetic inductances based on granular aluminum \cite{Maleeva, Gruenhaupt} or thin Nb nanowires \cite{niepce2019} will be suited. Also, the Josephson junctions should work in the charge regime $E_C \gtrsim E_J$, which guarantees a weak Landau level coupling.
In Table \ref{tab_current_circuit_values}, we list parameter values that are experimentally achievable in state-of-the-art superconducting circuits and that can be used to design GKP qubits. The resulting, relevant energy ratios which need to be small are $E_J / \hbar\omega_c = 0.4$ and $\hbar\omega_{LC} / E_J =0.8$. For the parameter defining the widths of the approximate grid states (see Sec. \ref{subsec_low_energy}), we obtain $\Delta=(E_L/E_J)^{1/4}e^{\pi/4}=0.8$.

\begin{table}[h]
   \centering
   \begin{tabular}{c|c}
   \textbf{Parameter} & $\mathrm{GHz}$ \\ 
   \hline
	$E_C/h $ 						& $13.50$ 	\\   
	$E_J/h$						& $3.50$		\\
	$E_L/h$						& $0.07$  		\\
	$\omega_{c}/2 \pi$		& $8.59$		\\
	$\omega_{LC}/2 \pi$		& $2.75$
   \end{tabular}
   \caption{Design parameters for the circuit in Fig. \ref{fig_circuit}. These parameters are achievable in state-of-the-art superconducting circuits. The charging energy and the inductive energy correspond to a capacitance $C=1.4\,\text{fF}$ and an inductance $L=2.3\,\mu\text{H}$, respectively.}
   \label{tab_current_circuit_values}
\end{table}

We emphasize that our circuit encodes the approximate grid states in a subsystem (to be precise, in the LLL) whose dynamics is effectively described by the approximate GKP Hamiltonian in Eq. \eqref{eq_H_parabolic_projected_with_x_p}. For this reason, the codewords are passively protected \cite{Doucot_protected}, and so, in contrast to current efforts to encode grid states in superconducting cavities \cite{Eickbusch}, they do not require permanent active stabilization \cite{Doucot_protected}.

We also point out that there is a different proposal for a superconducting circuit implementing grid states in a doubly non-linear qubit (the dualmon) \cite{Grimsmo}, which involves a Josephson junction and a quantum phase-slip wire. However, in contrast to our proposal, its dynamics is not described within a Landau level projection and also the GKP codewords are not the lowest-lying eigenstates of the resulting Hamiltonian \cite{Grimsmo}. \\

So far, we neglected the effect of potential external magnetic fluxes threading the superconducting loops in the circuit shown in Fig. \ref{fig_circuit}. Also, no external voltage or current sources have been attached to it. In the following, we show how these additional degrees of freedom can be used to perform single- and two-qubit gates and for state preparation and qubit read out. Because of the assumed symmetry between the fluxonia, the effective Hamiltonian is symmetric in $X$ and $P$ and so quantum operations can be performed in the logical $\overline{Z}$ or $\overline{X}$ basis depending on which port of the gyrator the sources are applied to.

\subsection{Logical $\overline{X}$ and $\overline{Z}$ gates}\label{subsec_x_z}
We now turn our attention to the implementation of logical gates in our system by focusing on the single-qubit $\overline{X}$ and $\overline{Z}$ gates defined in Eq. \eqref{eq_logical_X_Z}. For the sake of clarity, we will carry out the analysis for the case without harmonic confinement potential, i.e. without inductances in the superconducting circuit. The same procedures also work for the complete circuit in Fig. \ref{fig_circuit} when the ratios $E_L/E_J$ and $E_J/E_C$ are sufficiently small, yielding approximate logical gates. 

\begin{figure}[t]
	\centering
	\includegraphics[width=0.45\textwidth]{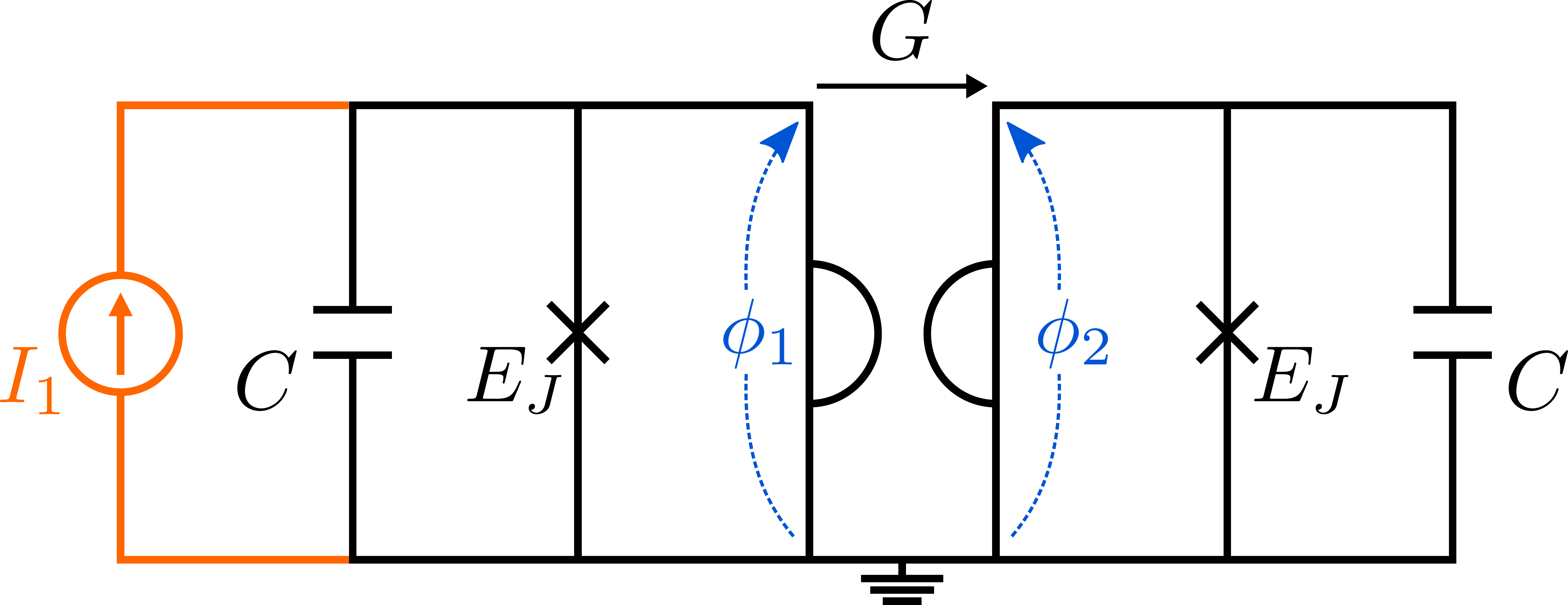}
	\caption{Circuit implementation of the \textit{ideal} logical $\overline{Z}$ gate with a current source. Assuming the system starts in the code subspace, the current source is turned on and kept constant at a value $I_1$ for a time $t_Z$ given in Eq. \eqref{eq_tz}. A current source on the opposite port of the gyrator would instead implement a logical $\overline{X}$ gate.}
	\label{fig_CJJG_Xgate}
\end{figure}

In our analysis, we demand that the system is operated in the relevant case of weak Landau level coupling as discussed in Sec. \ref{subsection_LLL_projection}. The logical operators $\overline{X}$ and $\overline{Z}$ can be implemented using current sources shunting either of the ports of the gyrator. The circuit implementing the $\overline{Z}$ gate is depicted in Fig. \ref{fig_CJJG_Xgate}. 

The Hamiltonian of this circuit can be written as
\begin{multline}\label{eq_circuit_Hamiltonian}
	H(t) = \frac{\bm{\pi}^2}{2 C} - E_J \cos \biggl[\frac{2 \pi}{\Phi_{0, s}}
			\biggl(R_1+	\frac{\pi_2}{C \omega_c}\biggr) \biggr] 					\\
			- E_J \cos \biggl[\frac{2 \pi}{\Phi_{0, s}} 
			\biggl(R_2-\frac{\pi_1}{C \omega_c}\biggr) \biggr]- I_1(t) 
			\biggl(R_1+ \frac{\pi_2}{C \omega_c} \biggr),
\end{multline}
where in analogy with the electronic case, see Eqs. \eqref{eq_def_dynamical_momenta} and \eqref{eq_def_guiding_center}, we defined the dynamical momenta $\pi_1= Q_1+G \phi_2/2$, $\pi_2= Q_2-G \phi_1/2$ and the guiding centers  $R_1= \phi_1-\pi_2/C \omega_c$, $R_2= \phi_2+\pi_1/C \omega_c$. The current source appears in the Hamiltonian in the term $-I_{1}(t) [R_1 + \pi_2/C \omega_c]=-I_{1}(t)\phi_1$. Note that in the electronic analogy, the current source acts as a homogeneous and time-dependent in-plane electric field, whose direction depends on the port the generator is connected to. We expect that if the current source $I_1(t)$ does not contain frequencies close to $\omega_c$, it will not cause transitions between Landau levels. Hence, we can project onto the LLL and obtain (dropping constant shifts in energy)
\begin{equation}\label{eq_H_LLL_with_I}
\begin{split}
	H_\text{LLL} 
	&= -V_0 \left[ \cos (2 \sqrt{\pi} X)+ \cos (2 \sqrt{\pi} P) \right] 
	- \frac{ I_1(t) \Phi_{0, s}}{\sqrt{\pi}} X  \\
	&= H_{\rm GKP}-  \frac{I_1(t) \Phi_{0, s}}{\sqrt{\pi}} X,
\end{split}
\end{equation}
where we immediately performed the variable rescaling in Eq. \eqref{eq_X_P_rescaling}. Also, we used the definition of the GKP Hamiltonian in Eq. \eqref{eq_GKP_Hamiltonian} to identify $V_0 = E_J e^{-\pi}$.
 
Now, we consider the following scenario. At time $t=0$, the state is assumed to be in a generic superposition $\ket{\psi_\text{in}}= c_0 \ket{\bar{0}}+c_1 \ket{\bar{1}}$ of the ideal GKP codewords given in Eq. \eqref{eq_GKP_code_words}. We assume a constant current source $I_1(t) \equiv I_1$. In this case, the time evolution operator associated with $H_\text{LLL}$ in Eq. \eqref{eq_H_LLL_with_I} reduces to
$U_\text{LLL}(t)= \exp(-i H_{\rm LLL} t/\hbar)$. In order to understand the effect of $U_\text{LLL}(t)$ on $\ket{\psi_\text{in}}$  we use the Zassenhaus formula \cite{Suzuki1977}
\begin{equation}\label{eq::zassenhaus}
	e^{t (A + B)} 
	= e^{t A} e^{t B} e^{-\frac{t^2}{2}[A, B]} e^{\frac{t^3}{6}(2 [B, [A, B]] + [A, [A, B]])} \dots
\end{equation} 
with $A=i I_1 \Phi_{0,s} X/\hbar \sqrt{\pi}$ and $B=-i H_{\rm GKP}/\hbar$. Since all the commutators in the Zassenhaus formula have the GKP states as degenerate eigenstates, e.g. 
\begin{equation}
[X, \cos(2 \sqrt{\pi} X) +  \cos(2 \sqrt{\pi} P)] = -i 2 \sqrt{\pi} \sin(2 \sqrt{\pi} P),
\end{equation}
one can show that
\begin{equation}
	\ket{\psi(t)} 
	= U_\text{LLL}(t) \ket{\psi_\text{in}}
	= e^{i \theta(t)} \exp(i \frac{I_1 \Phi_{0, s}}{\hbar \sqrt{\pi}} Xt) \ket{\psi_\text{in}},
\end{equation}
where we factorized the irrelevant phase factor $e^{i \theta(t)}$. Thus, after a time 
\begin{equation}\label{eq_tz}
	t_Z = \frac{\hbar \pi}{I_1 \Phi_{0, s}},
\end{equation}
a logical $\overline{Z}$ gate is applied to $\ket{\psi_\text{in}}$, up to an overall phase. Ideally, after a time $t_Z$, the current source must be switched off.
 
As already mentioned, because of the $X, P$ exchange symmetry of our circuit, it is  straightforward to convert between the logical $\overline{Z}$ and $\overline{X}$ basis by simply changing the port of the gyrator where the current source is applied and using the same protocol.

\subsection{Noise Sensitivity}
We provide a first analysis of the noise sensitivity of our qubit to typical noise sources, such as flux and charge noise. 
We start our discussion by analyzing charge noise. In the circuit in Fig. \ref{fig_circuit}, charge noise can be modeled by capacitively coupling random voltage sources to the ports of the gyrator. This modifications change the kinetic term in the Hamiltonian in Eq. \eqref{eq_LCJJ_G_LCJJ_Hamiltonian} as $\bm{\pi}^2/2C \mapsto (\bm{\pi} + \bm{Q}_g)^2/2C$, where $\bm{Q}_g$ is a vector containing the random charges on the capacitors connected to the voltage sources on either side of the circuit. The eigenspectrum is insensitive to static gate charges since they can be gauged away by a unitary transformation, as for the fluxonium qubit \cite{Manucharyan, ManucharyanPhD}. Moreover, charge noise couples to the dynamical momenta $\pi_{1, 2}$, and, as a consequence, it has only a small effect on the guiding center variables in which our states are encoded. From these arguments we conclude that charge noise should not be a major source of decoherence in our system, even if the transmon condition in not fulfilled \cite{KochTransmon}. 

Another typical noise source in our system is flux noise. We begin our analysis of flux noise sensitivity by considering again the ideal GKP Hamiltonian defined in Eq. \eqref{eq_GKP_Hamiltonian}, thus neglecting the confining potential of the inductive shunts. After the LLL projection, the external fluxes through the loops formed by gyrator branches and Josephson junctions give rise to the Hamiltonian \footnote{Note that we are here assuming that the gyrator forms a superconducting loop and, as a consequence, we are enforcing fluxoid quantization. Fluxoid quantization should not be enforced if the gyrator is either non-superconducting (e.g., the quantum Hall gyrator) or does not close a superconducting loop. These features will be dependent on the specific realization of the gyrator.}
\begin{equation}\label{eq_H_flux_noise}
\begin{split}
	H/V_0 = -\cos \left[ 2 \sqrt{\pi} X+\varphi_{G1}^\text{ext}(t) \right] 
					- \cos \left[2 \sqrt{\pi} P+\varphi_{G2}^\text{ext}(t) \right],
\end{split}
\end{equation}
where $\varphi_{G1,G2}^{\rm ext }(t)= 2 \pi \Phi_{G1,G2}^{\rm ext }(t)/\Phi_{0, s}$ are the reduced magnetic fluxes through the loops on either port of the gyrator, respectively. The GKP code space is intrinsically protected with respect to these noise sources as long as they are weak in strength. In order to show this, we rewrite Eq. \eqref{eq_H_flux_noise} as the sum of the desired GKP Hamiltonian and additional noise operators with time-dependent coefficients, i.e.
\begin{equation}\label{eq_H_flux_noise2}
\begin{split}
	H/V_0 = &H_{\rm GKP}/V_0 
	+ s_X(t)\sin(2 \sqrt{\pi}X) 
	+ c_X(t)\cos(2 \sqrt{\pi} X) \\
	&+ s_{P}(t)\sin(2 \sqrt{\pi}P)
	+ c_P(t) \cos(2 \sqrt{\pi} P),
\end{split}
\end{equation}
where we defined $c_{X, P}(t)= 1 - \cos[\varphi_{G1, G2}^\text{ext}(t)]$ and $s_{X, P}(t)= \sin[\varphi_{G1, G2}^\text{ext}(t)]$. Because all the individual noise operators in Eq. \eqref{eq_H_flux_noise2} have the GKP code space as degenerate eigensubspace, we conclude that the GKP code space is a decoherence-free subspace (DFS) \cite{Lidar, lidarDFS} with respect to this kind of noise. A similar observation was also made for the dualmon in Ref. \cite{Grimsmo}. We stress that this argument does not rely on the assumption of Markovianity of the flux noise.

The previous derivation assumed the ideal GKP Hamiltonian given in Eq. \eqref{eq_GKP_Hamiltonian}. However, since its spectrum is continuous and gapless, we want to work with its confined version \cite{Doucot_protected}, which corresponds to the circuit with inductances, shown in Fig. \ref{fig_circuit}. In this case, we have to take into account the noise associated with the external magnetic fluxes $\Phi_{1, 2}^\text{ext}$ through the superconducting loops formed by the Josephson junctions and the inductances on each port of the gyrator. We stress that in the limit of large inductances that we are considering here, this flux noise has a weak effect and that the associated noise term vanishes as the value of the inductances increases.

\begin{figure}
	\centering
	\includegraphics[width=0.40 \textwidth]{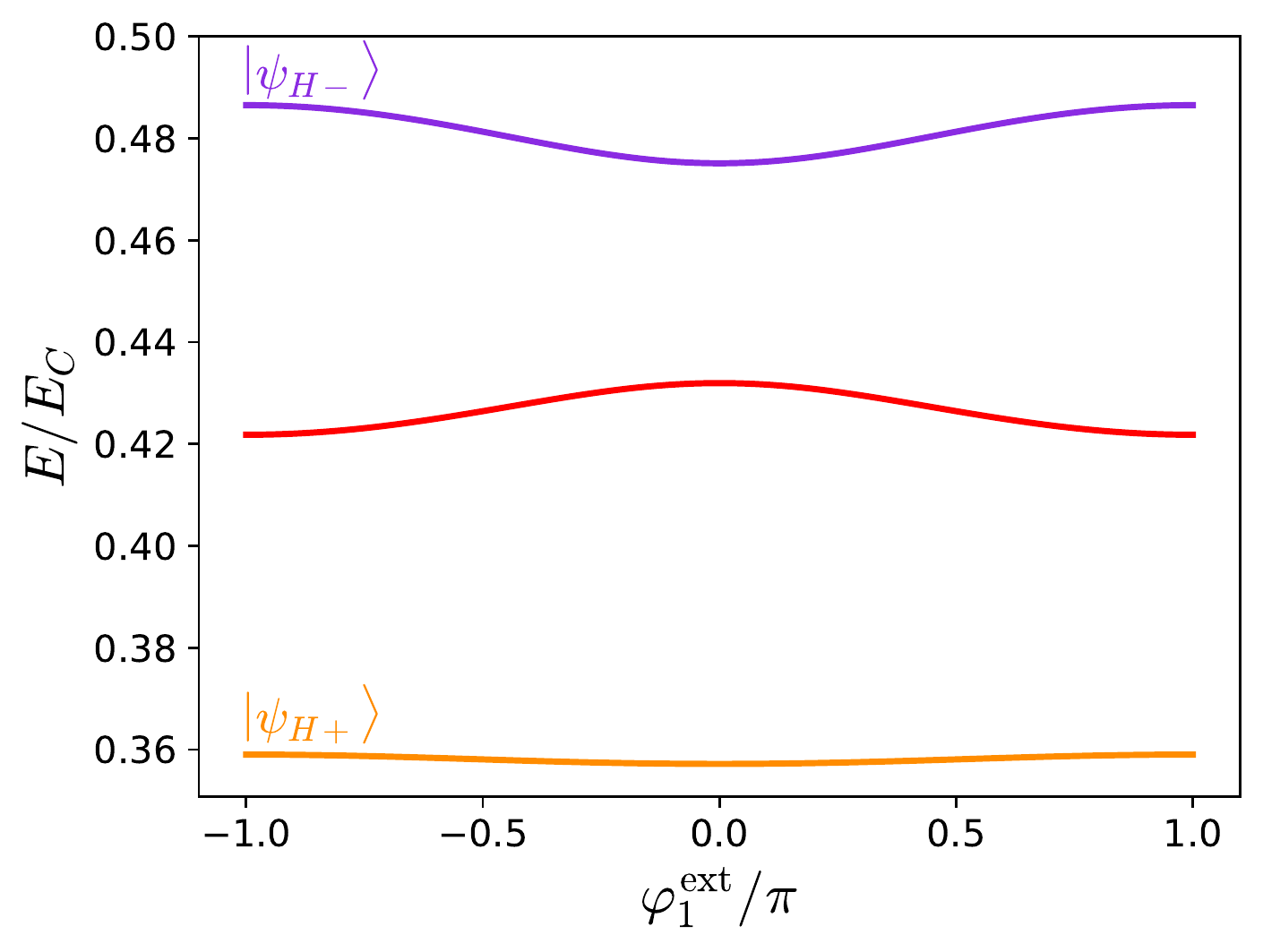} 
\caption{Low energy spectrum of the circuit shown in Fig. \ref{fig_circuit} as a function of the external flux $\varphi_{1}^{\rm ext} = 2 \pi \Phi_{1}^{\rm ext}/\Phi_{0, s}$ for fixed values of $E_J/E_C= 0.26$ and $E_L/E_C=5 \times 10^{-3}$ and with the other external fluxes set to zero. For these values, $\ket{\psi_{H-}}$ is the second excited state. One can clearly notice the sweet spots when $\varphi_{1}^{\rm ext}$ is an integer multiple of $\pi$.}
	\label{fig_dephFN}
\end{figure}

To avoid pure dephasing, the energy levels should not depend on the noise parameters. An example of the dependence of the energy levels on flux noise is shown in Fig. \ref{fig_dephFN}. Note that, like in the $0$-$\pi$ qubit \cite{Gyenis}, there is a level crossing of the second and third excited state as the inductances decrease, see also Fig. \ref{fig_lowest_10_energies}.
The protection against flux noise dephasing does not seem to be inherently different from that of state-of-the-art fluxonium qubits \cite{Manucharyan}, as well as the one of $0$-$\pi$ qubits \cite{Groszkowski} and bifluxon qubits \cite{Kalashnikov}, where the energy levels show a behavior as a function of the external fluxes similar to our qubit.

However, in analogy to the $0$-$\pi$ qubit \cite{Groszkowski} and the bifluxon \cite{Kalashnikov}, the disjoint support of the GKP codewords with respect to both $X$ and $P$ guarantees that the matrix elements of local noise operators between the encoded states are very small. This smallness, in turn, guarantees protection against energy relaxation. In our case, the relevant operators to characterize the noise due to the external fluxes are $\varphi_i = 2\pi \phi_i / \Phi_{0,s}$ (coupling to $\Phi^\text{ext}_i$ and $\Phi^\text{ext}_{Gi}$), and, to first order in the small noise parameters, $\sin(\varphi_i)$ (coupling to $\Phi^\text{ext}_{Gi}$). Furthermore, the noise operator associated with quasiparticle tunneling is $\sin(\varphi_i/2)$ \cite{Catelani}. These noise sources have a small effect in the relevant parameter regime. In fact, because the wave functions $\psi_{H\pm}(X)$ are both even in $X$, the matrix elements of the noise operators in the LLL projection between the two eigenstates $\ket{\psi_{H-}}$ and $\ket{\psi_{H+}}$ vanish, 
\begin{equation}
	\bra{\psi_{H-}} \Pi_{\text{LLL}} \mathcal{O}_\text{noise} \Pi_{\text{LLL}} \ket{\psi_{H+}}  = 0,
\end{equation} 
with $\mathcal{O}_\text{noise} \in \{\varphi_{1, 2},\, \sin(\varphi_{1,2}),\, \sin(\varphi_{1,2}/2)\}$.

\subsection{State Preparation, Qubit Readout and Clifford Gates}
State preparation, qubit readout and the implementation of Clifford gates are non-trivial and related topics for our qubit. A destructive measurement in the GKP basis $\{\ket{\bar{0}}, \ket{\bar{1}} \}$ can be performed by measuring the flux $\phi_1$, which in the LLL projection is approximately equivalent to measuring the rescaled guiding center variable $X$. An outcome of the measurement that is close to an even multiple of $\sqrt{\pi}$ corresponds to state $\ket{\bar{0}}$, while an outcome close to an odd multiple of $\sqrt{\pi}$ is assigned to state $\ket{\bar{1}}$.

A non-destructive measurement can be instead implemented if we have the ability to perform a GKP phase estimation protocol \cite{Weigand}, where we prepare an ancilla qubit in $\ket{\bar{0}}$, perform a CNOT with the ancilla qubit as target and then measure the ancilla destructively \cite{Deutsch1, Deutsch2}. In this protocol, the non-destructive measurement relies on the possibility to prepare the logical state $\ket{\bar{0}}$.

As recently shown in Ref. \cite{Yamasaki}, a logical $\ket{\bar{0}}$ state can be prepared deterministically by an adaptive protocol starting from two Hadamard eigenstates, using Clifford operations and a destructive readout of one of the two qubits. In Sec. \ref{subsec_low_energy}, we showed that the ground state of our system is indeed approximately a GKP Hadamard eigenstate. Thus, the ability to cool down our system in the ground state would give us also the ability to prepare the GKP $\ket{\bar{0}}$ state, when combined with Clifford operations and destructive measurement described above.
We also remark that in Ref. \cite{Menicucci}, it is shown instead how to prepare the GKP Hadamard eigenstate starting from many GKP logical $\ket{\bar{0}}$ states.

In Sec. \ref{subsec_x_z} we described a protocol to implement logical $\overline{X}$ and $\overline{Z}$ gates by means of current sources. Here we discuss further ideas for the implementation of general Clifford operations. As discussed in Ref. \cite{GKP}, one of the convenient properties of the GKP code is that, in the encoded subspace, Clifford unitaries are implemented by symplectic transformations (see also Ref. \cite{menicucciReview} for a review of gates for the GKP code). Symplectic transformations are generated by Gaussian unitaries and, as such, can be realized by using linear optics and squeezing \cite{Weedbrook, Adesso}. The Clifford group for a single qubit is generated by the Hadamard gate $\overline{H}$ defined in Eq. \eqref{eq_H_gate}, and the phase gate \cite{NielsenChuang}
\begin{equation}
	\overline{S} =
		\begin{pmatrix}
		1 & 0 \\ 
		0 & i
		\end{pmatrix}.
\end{equation}
As discussed in Ref. \cite{Doucot_protected} a possible way to implement the phase gate in our GKP qubits relies on the ability to change the magnitude of one of the quadratic terms in Eq. \eqref{eq_H_parabolic_projected_with_x_p}. This change can be achieved by tuning in time the superinductances, effectively creating an asymmetry between the two fluxonia in the circuit in Fig. \ref{fig_circuit}. In fact, in the GKP code the ideal unitary implementing the phase gate can be chosen as $U_{\overline{S}} = e^{-i X^2/2}$ \cite{GKP, menicucciReview} and so we need a term $\propto X^2$ that dominates the quadratic part of the Hamiltonian. This term appears if we create an asymmetry between the two inductances such that $E_{L_1} \gg E_{L_2}$. Then, in the LLL projection, we obtain a quadratic term $\propto E_{L_1} X^2$ that dominates over $\propto E_{L_2} P^2$. We note that the same effect can also be obtained by creating an asymmetry between the Josephson energies. This asymmetry can be achieved by substituting the Josephson junctions with SQUID loops \cite{SQUIDhandbook1, SQUIDhandbook2} and controlling the external fluxes in the loops. Similar ideas can also be employed to implement a CNOT gate: in this case, we need a tunable inductance coupling the branches of two of our GKP qubits. In addition, for the experimental realizable parameters in Table \ref{tab_current_circuit_values}, we believe that, in analogy to the $0$-$\pi$ qubit, one could perform gates also by using higher excitations of the circuit \cite{Paolo}. These ideas have been recently realized experimentally for the $0$-$\pi$ qubit \cite{Gyenis}.

The preparation of the ground state becomes more and more difficult as the quality of our GKP states improves. In fact, as the inductances increase, the ground and first excited states become closer in energy. In this case, state preparation would require temperatures that are lower than in current practice for superconducting qubits. 
Akin to the implementation of Clifford gates, an alternative approach to state preparation could use tunable superinductances. In this scheme, one prepares the ground state at relatively small inductances, and then adiabatically increases the inductances keeping the system always in the ground state, see Fig. \ref{fig_lowest_10_energies}. However, also this scheme becomes harder as the energy gap shrinks. As for the phase gate, we believe that a similar protocol can be realized by using SQUIDs instead of Josephson junctions and by modifying the effective Josephson energy by adiabatically tuning the external fluxes.

Here, we do not explore these protocols quantitatively, and leave a detailed description of the implementation to future research.
 
\section{Conclusions and Outlook}\label{sec_conclusions}
We have designed a circuit composed of state-of-the-art superconducting circuit elements and a non-reciprocal device, that can be used to passively implement the GKP quantum error correcting code. 
Our proposal crucially relies on the gyrator, which plays the role of an effective homogeneous magnetic field in an analogous electronic system and whose amplitude depends on the characteristic admittance of the device.
By taking advantage of recent advances in manufacturing non-reciprocal quantum Hall effect devices, one can reliably reach very high values of the effective magnetic field, which are well outside of the range that can be obtained in electronic systems.

By working out in detail the equivalence between our circuit and the problem of an electron in a magnetic field in a crystal potential, we analyze the system and identify a parameter range where the ground states of the system are the GKP codewords. Our analysis shows the deep relation between the GKP states and the Hofstadter butterfly, which, to the best of our knowledge, was not known previously.  
We study an implementation of approximate GKP codewords by shunting our circuit with large inductances.

We work out a mapping that allows to understand the eigenstates of the system in different coordinate systems, facilitating the interpretation of experimental results.

We discuss possible ways to implement one- and two-qubit logical gates as well as ideas for state preparation and qubit readout. This suggests that universal quantum computation can be done with our qubits by using only current sources and tunable inductances, or tunable Josephson junctions (SQUIDs). 

Finally, we discuss the effect of typical noise sources, i.e. charge and flux noise, and conclude that our qubit is well-protected against them. 

In this paper, we list a few ideas of how to implement phase gates and how to initialize the quantum state. A detailed comparison between the different protocols is still missing and is required to have a better understanding of the experimental capability of our qubit. Also, a more realistic modeling of the device would have to account for asymmetries in the circuit or for the internal degrees of freedom of the gyrator, whose effects have been overlooked in our analysis. However, we believe that these imperfections in the experiments would not affect the qualitative behavior of the system, which provides a promising hardware implementation of the GKP code.

\section*{Acknowledgments}
We gratefully acknowledge fruitful and continuous discussions with J. Conrad, F. Hassler, B. Terhal and D. Weigand. We also thank B. Terhal for carefully reading and commenting the manuscript. S. B. is supported by the Swiss National Science Foundation. A. C. is supported by ERC grant EQEC No. 682726. M. R. is funded by the Deutsche Forschungsgemeinschaft (DFG, German Research Foundation) under Germany's Excellence Strategy – Cluster of Excellence Matter and Light for Quantum Computing (ML4Q) EXC 2004/1 – 390534769.

\appendix

\section{Magnetic Translation Operators}\label{app_MTO}
In this appendix, we summarize a few key results about the magnetic translation operators (MTOs), which are required in Sec \ref{sec_crystal}. In analogy to the main text, here we restrict ourselves to the analysis of rational magnetic fluxes $\Phi/\Phi_0=p/q$ [see Eq. \eqref{eq_def_fraction}], where $\Phi=BL_0^2$ denotes the flux threading one unit cell of size $L_0 \times L_0$ and $\Phi_0=h/e$ is the magnetic flux quantum. Using the definition of the MTOs in Eq. \eqref{eq_MTO_def}, for integer values of $p$ and $q$ we find
\begin{equation}\label{eq_commuting_MTOs}
	[T_1(q L_0), T_2(L_0)] = 0,
\end{equation}
since the magnetic unit cell of size $q L_0 \times L_0$ contains $p$ flux quanta. Eq. \eqref{eq_commuting_MTOs} justifies the magnetic Bloch theorem in Eq. \eqref{eq_toroidal_BC}, which defines the Bloch states $\ket{\bm{k}}$, with $\bm{k}$ restricted to the first Brillouin zone, see Eq. \eqref{eq_Brillouin_zone}. 

Given the magnetic Bloch theorem in Eq. \eqref{eq_toroidal_BC}, we can find basis states that describe the system within a unit cell by considering the eigenvectors of the smallest possible translations compatible with the magnetic Bloch theorem. The choice of operators is of course non-unique, and here for example we choose the eigenvector of the operator $T_2(L_0/p)$,  which commutes with both $T_1(qL_0)$ and $T_2(L_0)$. We then define the basis states 
\begin{equation}\label{eq_MTO_appendix_label_l}
	T_2(L_0/p) \ket{\bm{k}, l} = e^{i (k_y L_0 + 2 \pi l)/p} \ket{\bm{k}, l},
\end{equation}
where $l=0,1,2,\ldots,p-1$.

Note that the Hamiltonian in Eq. \eqref{eq_final_crystal_Hamiltonian} exclusively comprises the MTOs $T_1(qL_0/p)$ and $T_2(qL_0/p)$ and their Hermitian conjugate. The action of $T_2(qL_0/p) = \left[ T_2 \left(L_0/p \right) \right]^q$ on $\ket{\bm{k}, l}$ follows straightforwardly from Eq. \eqref{eq_MTO_appendix_label_l}. Thus, it remains to analyze the action of $T_1(qL_0/p)$. From Eq. \eqref{eq_MTO_commutation}, we find
\begin{equation}
\begin{split}
	T_2 &\left(L_0/p \right) T_1 \left(qL_0/p \right) \ket{\bm{k}, l} 					\\
	&= e^{i (k_2 L_0 + 2 \pi (l+1) )/p} T_1 \left(qL_0/p \right) \ket{\bm{k}, l},
\end{split}
\end{equation}
from which we conclude that
\begin{equation}\label{eq_changing_l}
	T_1 \left(qL_0/p \right) \ket{\bm{k}, l} = e^{i k_1 q L_0 / p} \ket{\bm{k}, (l+1) \text{ mod } p}.
\end{equation}
Note that the state $\ket{\bm{k}, l}$ maps into itself after $p$ consequent applications of $T_1 \left(qL_0/p\right)$, in agreement with Eq. \eqref{eq_toroidal_BC}.

Finally, we analyze the degeneracy of the eigenstates of the Hamiltonian $H$ in Eq. \eqref{eq_final_crystal_Hamiltonian}. To this end, we note that $T_1(L_0)$ commutes with $H$ but not with both the MTOs of the boundary conditions in Eq. \eqref{eq_toroidal_BC}. Thus, if $\ket{\psi}$ is an eigenstate of $H$ with eigenenergy $E$, the state $T_1(L_0) \ket{\psi}$ is also an eigenstate of $H$ with the same eigenenergy, and because $[T_1(L_0), T_2(L_0)] \neq 0$, the states $\ket{\psi}$ and $T_1(L_0) \ket{\psi}$ are physically distinguishable for $q>1$. In particular, one can easily show that
\begin{equation}
\begin{split}
	T_1(L_0) &\ket{(k_1, k_2)^T, l} \\
	&= e^{i k_1 L_0} \ket{[k_1, (k_2 + 2\pi / q L_0) \text{ mod } 2 \pi / L_0]^T, l}	 \\
	&\not\propto \ket{(k_1, k_2)^T, l}	,	\qquad \text{for } q>1.
\end{split}
\end{equation}
As a result, every energy-band of the Hamiltonian is at least $q-$fold degenerate.

\section{Numerical Analysis of the Eigensystem}
\subsection{Without Confinement Potential}\label{app_numerics_crystal_electron}
In the following, we provide a method to numerically determine the spectrum of the Hamiltonian in Eq. \eqref{eq_final_crystal_Hamiltonian}. To this end, we expand the Hamiltonian in the product state basis $\ket{n;\bm{k},l}$ [defined in Eqs. \eqref{eq_def_Landau_Level_Fock_state} and \eqref{eq_toroidal_2}] with $l=0,\ldots,p-1$ and $n=0,\ldots,N$ for some reasonably large integer $N$.

In particular, the matrix elements of the displacement operator in the Landau level basis are known analytically \cite{Glauber2} and read
\begin{equation}\label{eq_Glauber}
	\bra{m} D_a(\alpha) \ket{n} = 
	\sqrt{\frac{n!}{m!}} \alpha^{m-n} e^{-|\alpha|^2/2} L_n^{m-n}(|\alpha|^2),
\end{equation}
where $L_n^{m-n}(|\alpha|^2)$ denotes the associated Laguerre polynomial.
Note that the evaluation of the Hamiltonian in the given basis is particularly convenient in the weak Landau level coupling limit ($V / \hbar \omega_c \ll 1$), since the coupling of product states $\ket{n;\bm{k},l}$ with different $n$ is weak. In this limit, every Landau level splits into $p$ bands with finite widths, which are well separated from all the other split Landau levels, see Fig. \ref{fig_energy_spectrum}. Considering only one Landau level, it is worth mentioning that the way in which it splits results in an energy spectrum which shows a fractal behavior similar to a deformed Hofstadter butterfly \cite{Hofstadter}, see Fig. \ref{fig_Hofstadter_butterfly}.

\subsection{Including Confinement Potential}\label{app_numerical_preparation}
Here, we present a convenient basis for the numerical analysis of the eigensystem of the Hamiltonian in Eq. \eqref{eq_confined_Hamiltonian_initial}. To this end, we introduce the bosonic ladder operators associated with the guiding center variables,
\begin{equation}
	b =  \frac{1}{\sqrt{2}}\frac{R_1 + i R_2}{l_B},		\qquad
	b^\dagger =  \frac{1}{\sqrt{2}}\frac{R_1- i R_2}{l_B},
\end{equation}
satisfying $[b,b^\dagger]=1$, and define the unitary displacement operator associated to these variables,
\begin{equation}
	D_b(\beta)=e^{\beta b^\dagger - \beta^* b},	\qquad \beta \in \mathbb{C}.
\end{equation}
Given the ladder operators of the guiding center variables and those of the dynamical momenta [see Eq. \eqref{eq_ladder_momenta}], we rewrite the Hamiltonian in Eq. \eqref{eq_confined_Hamiltonian_initial} as (dropping constant energy offsets)
\begin{equation}\label{eq_confined_Hamiltonian}
\begin{split}
	H &=  \hbar \omega_c a^\dagger a + \frac{\hbar\omega_0^2}{\omega_c} 
	\left( a^\dagger a + b^\dagger b + a b + a^\dagger b^\dagger \right) 				\\
	&\quad - \frac{V}{2} \bigg[ D_a(\lambda) D_b(-\lambda) + D_a(i \lambda) D_b(i \lambda)	+ \mathrm{h.c.} \bigg],
\end{split}
\end{equation}
with $\lambda = \sqrt{q\pi/p}$ being the absolute value of each displacement. This Hamiltonian will be expanded in the basis of the Fock product-states
\begin{equation}
	\ket{n,m} 
	= \frac{ {a^\dagger}^n}{\sqrt{n!}} \ket{0} \otimes \frac{ {b^\dagger}^m}{\sqrt{m!}} \ket{0}
	, \qquad n, m  \in \mathbb{N}_0,
\end{equation}
whereat we have to reasonably truncate $n$ and $m$. In the process, the matrix elements of each individual term in the Hamiltonian are known analytically, especially the matrix elements of the displacement operators, see Eq. \eqref{eq_Glauber}. \\

At this point, one could proceed with an analytical diagonalization of the quadratic part \cite{Xiao} of the Hamiltonian [first line in Eq. \eqref{eq_confined_Hamiltonian}] in order to reduce the coupling of the basis states. The eigenstates of the quadratic part of the Hamiltonian are known as Fock-Darwin states \cite{Fock, Darwin}.

We, however, do not perform this diagonalization because we want to retain the jargon of Landau levels. Nevertheless, both approaches coincide in the limit of consideration. \\

After the LLL projection, i.e. restricting to the subspace spanned by $\ket{0,m}$, the Hamiltonian in Eq. \eqref{eq_confined_Hamiltonian} reduces to (dropping constant energy offsets)
\begin{equation}\label{eq_H_parabolic_projected}
	H_\text{LLL} = \frac{\hbar\omega_0^2}{\omega_c} b^\dagger b 
	- \frac{V_0}{2} \bigg[ D_b(i \lambda) + D_b(-i \lambda) + D_b(\lambda) + D_b(-\lambda) \bigg]
\end{equation}
where $V_0 = V e^{-\pi q / 2p}$. Note that in the limit of weak confinements ($\hbar\omega_0^2/\omega_cV_0 \ll 1$), states $\ket{0,m}$ with different $m$ are strongly coupled due to the crystal potential. Therefore, a large number of Fock states is required for an accurate numerical treatment. Nevertheless, a finite confinement prevents the eigenstates of constituting arbitrarily high excited Fock states.

Moreover, the matrix elements $\bra{0, m_1} H_\text{LLL} \ket{0, m_2}$, are non-zero only if $m_1 = m_2 \! \mod 4$. Thus, the Hamiltonian in the LLL projection couples only every fourth Fock state. For this reason, also the eigenstates of $H_\text{LLL}$ comprise only every fourth Fock state \cite{GKP}. 

The consequence of this characteristic of the eigenstates becomes clear by considering the $X$-representation [see Eq. \eqref{eq_X_P_rescaling}] of the $m$'th Fock state,
\begin{equation}\label{eq_Hermite_functions}
	\braket{X}{m} = \frac{1}{\sqrt[4]{\pi}} \frac{1}{\sqrt{2^m m!}} H_m(X) e^{-X^2/2},
\end{equation}
where $H_m(X)$ is the the $m$'th Hermite polynomial. The Hermite functions in Eq. \eqref{eq_Hermite_functions} have the fundamental property of being eigenfunctions of the Fourier transform with eigenvalue $(-i)^m$ \cite{Husimi_Fock}, which is cyclic in $m$ with periodicity 4. Hence, we can conclude that also the eigenfunctions of the effective Hamiltonian in Eq. \eqref{eq_H_parabolic_projected} are invariant under a Fourier transform, up to a constant prefactor $(-i)^m$. 

\section{Envelope Function Approximation - Derivation of the Approximate Grid States}\label{appendix_envelope_function_theory}
In the following, we derive the approximate grid states introduced in Sec. \ref{subsec_low_energy}, by using the envelope function approximation. For convenience, we rescale the Hamiltonian in Eq. \eqref{eq_H_parabolic_projected_with_x_p} by $\hbar\omega_0^2/\omega_c$, leading to 
\begin{equation}\label{eq_H_for_envelope_function_approx}
	H = \frac{P^2+X^2}{2} - W \left[ \cos(2\sqrt{\pi}X) + \cos(2\sqrt{\pi}P)  \right],
\end{equation}
where we introduce the large dimensionless parameter $W = \omega_c V_0 / \hbar \omega_0^2 \gg 1$. \\

Let us first examine the symmetries of this Hamiltonian.

First, $H$ is an even function of both $X$ and $P$, and so in both $X$- and $P$-representation, its eigenfunctions can be chosen to be even and odd real-valued functions.

Second, since the Hamiltonian is invariant under an exchange of the position and momentum variables ($X \leftrightarrow P$), its eigenfunctions must be equal (up to an overall phase) in both representations. From this statement, it follows that the non-degenerate eigenfunctions of the Hamiltonian must be eigenfunctions of the Fourier transform.

In general, applying the Fourier transform twice is equivalent to applying a parity operation; applying the Fourier transform four times corresponds to the identity. For this reason, the eigenvalues of the Fourier transform are integer powers of $i$. In particular, the eigenfunctions of the Fourier transform with even parity have eigenvalues $\pm 1$ under Fourier transform. \\

After having analyzed the Hamiltonian's symmetries, we now determine its approximate eigenfunctions with a consequent double application of the envelope function approximation \cite{Girvin, Kittel, Rossi}.

Because $W \gg 1$, we start from the GKP Hamiltonian
\begin{equation}
	H_\text{GKP} =  - W \left[ \cos(2\sqrt{\pi}X) + \cos(2\sqrt{\pi}P)  \right],
\end{equation}
and we treat the extra terms as smooth perturbations, determining the behavior of the envelope function that modulates the GKP ground states. The eigenstates of $H_\text{GKP}$ are uniquely defined by the Zak states \cite{Zak, Zak2, Zak3} \footnote{Comparing with Eq. \eqref{eq_toroidal_2} in which we set $p/q=1/2$, note that the boundary conditions in Eq. \eqref{eq_Zak_BC} correspond to the opposite enlargement of the unit cell.}
\begin{subequations}\label{eq_Zak_BC}
\begin{align}
	e^{i 2\sqrt{\pi} X} \ket{\Psi_{k,q}} &= e^{i 2\sqrt{\pi} q} \ket{\Psi_{k,q}}, 		\\
	e^{i \sqrt{\pi} P} \ket{\Psi_{k,q}} &= e^{i \sqrt{\pi} k} \ket{\Psi_{k,q}}, 
\end{align}
\end{subequations}
with $k \in [-\sqrt{\pi}/2, 3\sqrt{\pi}/2)$ and $q \in [-\sqrt{\pi}/2, \sqrt{\pi}/2)$. In position representation, these states can be written in the Bloch form 
\begin{equation}\label{eq_Zak_Bloch_form}
	\Psi_{k, q}(X) = e^{ikX} \sum_{n \in \mathbb{Z}} \delta(X - \sqrt{\pi} n - q),
\end{equation}
and, by a Fourier transform, we obtain (up to a global prefactor)
\begin{equation}\label{eq_Zak_Fourier_transform}
	\Psi_{k, q}(P) = e^{-iqP} \sum_{n \in \mathbb{Z}} \delta(P - 2\sqrt{\pi} n - k).
\end{equation}
The spectrum of $H_\text{GKP}$ is continuous and its bandstructure is given by
\begin{equation}
	E(k, q) = - W \bigg[ \cos(2\sqrt{\pi}k) + \cos(2\sqrt{\pi}q)  \bigg].
\end{equation}
This band has two degenerate minima with energy $E=-2W$, obtained for $(k=0, q=0)$ and $(k=\sqrt{\pi}, q=0)$. \\

Let us now consider the Hamiltonian $H_P=H_{GKP}+P^2/2$. Because the latter term is smooth on the scale of the GKP Hamiltonian, we assume that the eigenfunction of $H_P$ can be factorized as
\begin{equation}
\label{eq:h1}
\widetilde{\Psi}_{k,q}(P)=\phi(P)\Psi_{k,q}(P),
\end{equation} 
where the Bloch function $\Psi_{k,q}(P)$ is given in Eq. \eqref{eq_Zak_Fourier_transform}. The function $\phi(P)$ is a smooth envelope that modulates the Bloch function and in analogy to solid-state theory, it satisfies \cite{Girvin, Kittel, Rossi}
\begin{equation}
\label{eq_first_envelope}
	\left( E(k,q) + \frac{P^2}{2} \right)\phi(P) = \mathcal{E}(k) \phi(P),
\end{equation}
with the eigenvalue $\mathcal{E}(k)$. We are interested in the ground state eigenfunctions only, and so we expand $E(k,q)$ around the minimum $q=0$ (effective mass approximation). Neglecting a constant energy offset and promoting the crystal momentum $q$ to the operator $i\partial_P$, we obtain the harmonic oscillator differential equation
\begin{equation}
	\left( -2 \pi W \partial_P^2 + \frac{P^2}{2} \right) \phi(P)
	= \left[ \mathcal{E}(k) + W \cos(2 \sqrt{\pi} k) \right] \phi(P) ,
\end{equation}
which has the ground state wave function
\begin{equation}
	\phi(P) = \frac{\sqrt{\Delta}}{\pi^{1/4}} \exp(-\frac{\Delta^2 P^2}{2}).
\end{equation}
 The characteristic length of this oscillator is 
\begin{equation}\label{eq_typical_length}
	1/\Delta = \left( 4 \pi W \right)^{1/4}
\end{equation}
and corresponds to the broadening $1/\Delta$ of the Gaussian envelope function discussed in Sec. \ref{subsec_low_energy}, see Eq. \eqref{eq_width_delta}. \\

To include the $X^2/2$ term, we first Fourier transform Eq. \eqref{eq:h1} for $q=0$, leading to the Bloch functions in $X$,
\begin{equation}
\widetilde{\Psi}_{k,0}(X)\approx e^{i k X} \sum_{n \in \mathbb{Z}} e^{-(X-\sqrt{\pi}n+k)^2/2\Delta^2},
\end{equation}
where the approximate sign holds in the limit  $\Delta\ll 1$.
Because $X^2/2$ varies smoothly in each period of the Bloch function, we proceed as before and factorize the wave function of $H = H_P + X^2/2$ as
\begin{equation}
\psi(X) =\Phi(X)\widetilde{\Psi}_{k,0}(X),
\end{equation}
where the modulating function $\Phi(X)$ satisfies the eigenvalue equation
\begin{equation}\label{eq_second_envelope}
	\left(\mathcal{E}(k) + \frac{X^2}{2} \right)\Phi(X) = \epsilon \, \Phi(X),
\end{equation}
and in the effective mass approximation, is given by
\begin{equation}
	\Phi(X) = \frac{\sqrt{\Delta}}{\pi^{1/4} } \exp(-\frac{\Delta^2X^2}{2 }).
\end{equation}
Importantly, $\mathcal{E}(k)$ has two degenerate minima at $k=0$ and at $k=\sqrt{\pi}$, and so we obtain two approximate ground state eigenfunctions, that are given by the broadened GKP codewords $\psi_0(X)$ and $\psi_1(X)$ defined in Eq. \eqref{eq_fit_approx_grid_states}. Note that these states are approximately orthonormal in the limit $\Delta \ll 1$. Importantly, in the approximation used here, the eigenstates are degenerate and the Fourier transforms of $\psi_{0,1}(X)$ are approximately given by
\begin{equation}
	\psi_\pm(X)=\frac{\psi_0(X)\pm\psi_1(X)}{\sqrt{2}}.
\end{equation}
Thus, we construct the states
\begin{subequations}
\begin{align}
\psi_{H+}(X) &= \cos( \frac{\pi}{8} ) \psi_0(X) + \sin( \frac{\pi}{8} ) \psi_1(X),		\\
\psi_{H-}(X) &= -\sin( \frac{\pi}{8} ) \psi_0(X) + \cos( \frac{\pi}{8} ) \psi_1(X),
\end{align}
\end{subequations}
that are approximately even and odd functions under Fourier transform, and respect the symmetries of the Hamiltonian. \\
 
As a consistency check, we can also estimate the broadening of the codewords in Fig. \ref{fig_approx_grid_states}, by using the Heisenberg uncertainty principle.
For weak confinements $\hbar\omega_0^2/\omega_c V_0\ll 1$, we expect the low-energy eigenfunctions of the Hamiltonian in Eq. \eqref{eq_H_parabolic_projected_with_x_p} to have support only in the vicinity of $X=n 2\sqrt{\pi}$ and $X=n 2\sqrt{\pi}+\sqrt{\pi}$, respectively, with $n \in \mathbb{Z}$.

Let us focus on what happens close to $X=0$.
From Heisenberg's uncertainty principle, confining a particle to a narrow region causes large fluctuations in  momentum $P$, such that $\hbar\omega_0^2 \expval{P^2}/\omega_c \gg V_0$. By expanding $\cos(2\sqrt{\pi}X)$ up to second order and neglecting the fast oscillating term $\cos(2\sqrt{\pi}P)$ and the small perturbation $\hbar\omega_0^2 X^2/\omega_c$, we obtain
\begin{equation}
	H \approx \frac{\hbar \omega_0^2 P^2}{2\omega_c} + 2 \pi V_0 X^2.
\end{equation}
This Hamiltonian is the one of a harmonic oscillator and its ground state is a Gaussian of width 
\begin{equation}
	\Delta = \sqrt[4]{\frac{\hbar \omega_0^2}{4 \pi \omega_c V_0}}.
\end{equation} 
This wave function approximates the narrow Gaussian centered at $X=0$ of the approximate GKP state $\psi_0(X)$. 

The width of the wide Gaussian envelope can be found by a similar argument, where one first considers localization in momentum $P$, and then Fourier transforms the result, leading to Eq. \eqref{eq_typical_length}.

\section{Derivation of the Integration Kernel}\label{app_kernel}
In this appendix, we derive the analytical expression of the integration kernel in Eq. \eqref{eq_integration_kernel}, connecting the wave functions of the two-dimensional Hamiltonians and that of the one-dimensional Hamiltonians projected onto the LLL. 
By inserting the identity operator
\begin{equation}
	\widehat{\bm{1}}
	= \sum_{n=0}^\infty \int_{-\infty}^\infty \! dX \ketbra{n, X}{n, X},
\end{equation}
in $\Psi(x_1, x_2) = \braket{x_1, x_2}{\Psi}$, and neglecting mixing to higher Landau levels, we find
\begin{equation}\label{eq_def_kernel}
	K_0(x_1, x_2 ; X) = \braket{x_1, x_2}{0, X}.
\end{equation}
To derive Eq. \eqref{eq_integration_kernel}, we introduce the annihilation operators of the dynamical momenta and guiding center variables,
\begin{equation}
	\widehat{a} = \frac{1}{\sqrt{2}}\frac{l_B}{\hbar} \left( \widehat{\pi}_2 + i \widehat{\pi}_1 \right),
	\qquad
	\widehat{b} = \frac{1}{\sqrt{2}}\frac{1}{l_B} \left( \widehat{R}_1 + i \widehat{R}_2 \right).
\end{equation}
Note that, for the sake of clarity, within this appendix, we indicate operators with hats on top. Also, from now on, the coordinates $x_i$ are given in magnetic units, thus scaled by the magnetic length $l_B$. Inserting the definition of the guiding center variables $\widehat{R}_i$ [cf. Eq. \eqref{eq_def_guiding_center}] in this equation yields 
\begin{equation}
	\widehat{b} = \frac{\widehat{x}_1 + i \widehat{x}_2}{\sqrt{2}} - \widehat{a}^\dagger.
\end{equation}
The projector $\widehat{\Pi}_\text{LLL} = \ketbra{0}_a$ onto the LLL acts solely in the Hilbert space of the dynamical momenta and thus commutes with any operator acting exclusively on the guiding center variables. Because $[ \widehat{\Pi}_\text{LLL}, \widehat{b} ]=0$, we find that the action of $\widehat{b}$ on the projected position eigenstate,
\begin{equation}\label{eq_coherent_state}
	\widehat{b} \, \widehat{\Pi}_\text{LLL} \ket{x_1, x_2}
	= \frac{x_1 + i x_2}{\sqrt{2}} \widehat{\Pi}_\text{LLL} \ket{x_1, x_2},
\end{equation}
is equivalent to the definition of a coherent state \cite{Gerry} $\ket{\beta}_b$ in the subspace of the guiding center variables, i.e.
\begin{equation}
	\widehat{\Pi}_\text{LLL} \ket{x_1, x_2} \propto \ket{0}_a \otimes \ket{\beta = (x_1 + i x_2)/\sqrt{2}}_b,
\end{equation}
We stress that Eq. \eqref{eq_coherent_state} is only valid for projectors onto the \textit{lowest} Landau level, since $\widehat{\Pi}_\text{LLL} \widehat{a}^\dagger= 0$. \\

In conclusion, the integration kernel as defined in Eq. \eqref{eq_def_kernel} is the complex conjugate of the coherent state wave function in position representation \cite{Gerry},
\begin{equation}\label{eq_Kernel_with_g}
\begin{split}
	K_0(x_1, x_2 ; X)
	= \frac{1}{\sqrt{2}\pi^{3/4}} \exp( - \frac{(X - x_1)^2}{2} ) \\ 
	\times \exp(-i x_2 X) \exp( i g(x_1, x_2) ),
\end{split}
\end{equation}
where the prefactor is chosen to satisfy
\begin{equation}
	\iint \! dx_1 dx_2 \, K^*_0(x_1, x_2; X) K_0(x_1, x_2; X') = \delta(X-X').
\end{equation}
The real-valued function $g(x_1, x_2)$ arranges an adjustment of the global complex phase for fixed values of $x_i$. It is determined by the chosen gauge of the vector potential $\bm{A}(x_1, x_2)$ and is fixed by demanding the state $\ket{{0, X}}$ to lie in the LLL. Since the annihilation operator $\widehat{a}$ of the dynamical momenta, expressed in the initial positions $\widehat{x}_i$, is gauge dependent, the integration kernel has to satisfy (in magnetic units)
\begin{equation}
\begin{split}
	0 
	=
	\hspace{-3pt}
	\bigg[
	\partial_{x_1}	
	-i \partial_{x_2} 
	+ A_2(x_1, x_2)
	+ i A_1(x_1, x_2)				
	\bigg]
	K_0(x_1, x_2; X),
\end{split}
\end{equation}
leading to the equation
\begin{equation}
	- \grad g(x_1, x_2) =
	\begin{pmatrix}
		A_1(x_1, x_2) \\
		A_2(x_1, x_2) - x_1
	\end{pmatrix}
\end{equation}
for the function $g(x_1, x_2)$. In particular, for symmetric gauge (in magnetic units),
\begin{equation}
	A_1(x_1, x_2) = -\frac{x_2}{2},	\qquad
	A_2(x_1, x_2) = \frac{x_1}{2},
\end{equation}
we find, up to a trivial constant,
\begin{equation}\label{eq_gauge_function}
	g(x_1, x_2) = \frac{x_1 x_2}{2}.
\end{equation}
Combining Eqs. \eqref{eq_Kernel_with_g} and \eqref{eq_gauge_function}, we obtain Eq. \eqref{eq_integration_kernel}.


\section{Relation between the four-fold Rotation and the Fourier Transform}\label{app_4fold_rotation}
Given the integral transform in Eq. \eqref{eq_integral_transform}, here, we want to show that any one-dimensional eigenfunction of the Fourier transform with eigenvalue $(-i)^n$, i.e.
\begin{equation}
	\frac{1}{\sqrt{2\pi}} \int_{-\infty}^\infty \! dP \psi(P) e^{-iXP} = (-i)^n\psi(X),
\end{equation}
results in a two-dimensional eigenfunction of the four-fold rotation with the complex conjugate eigenvalue. To this end, we make use of the result for the integration kernel in symmetric gauge [see Eq. \eqref{eq_integration_kernel}], and obtain
\begin{equation}
\begin{split}
	&\Psi(x_2, -x_1) 																								\\
	&= \int_{-\infty}^\infty \! dX K_0(x_2, -x_1 ; X) \psi(X) 										\\
	&= \frac{i^n}{2\pi^{7/4}} \iint_{\mathbb{R}^2} \! dX dP 
		e^{- \frac{(X - x_2)^2}{2}} e^{i (x_1 - P) X} e^{- \frac{x_1 x_2}{2}} \psi(P) 		\\	
	&= \frac{i^n}{\sqrt{2}\pi^{3/4}} \int_{-\infty}^\infty \! dP
		e^{-\frac{(P-x_1)^2}{2}} e^{-ix_2P} e^{i \frac{x_1x_2}{2}} \psi(P) 					\\	
	&= i^n \int_{-\infty}^\infty \! dP K_0(x_1, x_2 ; P) \psi(P) 									\\
	&= i^n \Psi(x_1, x_2).
\end{split}
\end{equation}
The particular choice of the symmetric gauge is essential for the previous derivation, since it determines the complex phase of the integration kernel for fixed values of $x_i$. This is in agreement with the observation the two-dimensional Hamiltonian in Eq. \eqref{eq_confined_Hamiltonian_initial} is four-fold rotational symmetric in the $x_1x_2$-plane in the symmetric gauge only.

\bibliography{bib_file.bib}

\end{document}